\documentclass[a4paper,prl,reprint,superscriptaddress,twocolumns,notitlepage,floatfix]{revtex4-1}
\usepackage[utf8]{inputenc}

\usepackage{graphicx}
\usepackage{amsmath}
\usepackage{mathrsfs}  
\usepackage{braket}
\usepackage{amssymb} 
\usepackage{csquotes}
\usepackage{array}
\usepackage{booktabs}
\usepackage{multirow}
\usepackage{xcolor}
\usepackage{siunitx}
\usepackage{pifont}
\usepackage{microtype}

\RequirePackage{expl3,xparse,l3regex}
\ExplSyntaxOn
\NewDocumentCommand{\hm}{m}
{
    \tl_set:Nx \l_hm_contents {\ensuremath{#1}}
    \regex_replace_all:nnN { -(.) } { \c{overline}\cB\{ \1 \cE\} } \l_hm_contents
    \regex_replace_all:nnN { (.)\_(.) } { \1\c{c_math_subscript_token}\2 } \l_hm_contents
    \regex_replace_all:nnN { ([a-zA-Z]+) } { \c{text}\cB\{ \1 \cE\} } \l_hm_contents
    \tl_use:N \l_hm_contents
}

\NewDocumentCommand{\sch}{m}
{
    \tl_set:Nx \l_sch_contents {#1}
    \regex_replace_all:nnN { \_ }{} \l_sch_contents \regex_replace_all:nnN { (.)(.*) } { \1\c{c_math_subscript_token}\cB\{\2\cE\} } \l_sch_contents \regex_replace_all:nnN { (d|h|v|s) }{ \c{text}\cB\{\1\cE\} }  \l_sch_contents \regex_replace_all:nnN { \cS\  }{} \l_sch_contents \tl_set:Nn \l_sch_result {\ensuremath{\l_sch_contents}} \tl_use:N \l_sch_result
}

\NewDocumentCommand{\irrep}{m}
{
    \tl_set:Nx \l_sch_contents {#1}
    \regex_replace_all:nnN { \_ }{} \l_sch_contents \regex_replace_all:nnN { (.)(.*) } { \1\c{c_math_subscript_token}\cB\{\2\cE\} } \l_sch_contents \regex_replace_all:nnN { ([a-zA-Z]) }{ \c{text}\cB\{\1\cE\} }  \l_sch_contents \regex_replace_all:nnN { \cS\  }{} \l_sch_contents \tl_set:Nn \l_sch_result {\ensuremath{\l_sch_contents}} \tl_use:N \l_sch_result
}
\ExplSyntaxOff

\DeclareFontFamily{U}{crystallographicSymbol}{\hyphenchar\font=-1}
\DeclareFontShape{U}{crystallographicSymbol}{m}{n}{ <-> cryst}{}
 
\usepackage{hyperref}
\usepackage{bookmark}

\definecolor{tab_blue}{HTML}{1F77B4}
\definecolor{tab_orange}{HTML}{FF7F0E}
\definecolor{tab_green}{HTML}{2CA02C}
\definecolor{tab_red}{HTML}{D62728}
\definecolor{tab_purple}{HTML}{9467BD}
\definecolor{tab_brown}{HTML}{8C564B}
\definecolor{tab_pink}{HTML}{E377C2}
\definecolor{tab_gray}{HTML}{7F7F7F}
\definecolor{tab_olive}{HTML}{BCBD22}
\definecolor{tab_cyan}{HTML}{17BECF}

\hypersetup{breaklinks=true,
  colorlinks=true,
  linkcolor=tab_blue,
  filecolor=tab_blue,
  urlcolor=tab_blue,
  citecolor=tab_blue,
}

\def\Id{\text{Id}}

\def\tr{\ensuremath{\operatorname{tr}}}

\def\diag{\ensuremath{\operatorname{diag}}}
\def\ee{{\rm e}}
\def\ii{{\rm i}}

\def\RR{\ensuremath{\mathbb{R}}}

\def\strong#1{\textbf{#1}}

\AtBeginDocument{
\heavyrulewidth=.08em
\lightrulewidth=.05em
\cmidrulewidth=.03em
\belowrulesep=.65ex
\belowbottomsep=0pt
\aboverulesep=.4ex
\abovetopsep=0pt
\cmidrulesep=\doublerulesep
\cmidrulekern=.5em
\defaultaddspace=.5em
}

\def\paragraphtitle#1{\emph{#1} --} 
\makeatletter\begin{document}

\title{Simple views on symmetries and dualities in the theory of elasticity}
\author{Michel Fruchart}
\email{fruchart@uchicago.edu}
\affiliation{James Franck Institute and Department of Physics, University of Chicago, Chicago IL 60637, USA}
\author{Vincenzo Vitelli}
\email{vitelli@uchicago.edu}
\affiliation{James Franck Institute and Department of Physics, University of Chicago, Chicago IL 60637, USA}
\date{\today} 

\begin{abstract}
Microscopic symmetries impose strong constraints on the elasticity of a crystalline solid. 
In addition to the usual spatial symmetries captured by the tensorial character of the elastic tensor, hidden non-spatial symmetries can occur microscopically in special classes of mechanical structures. 
Examples of such non-spatial symmetries occur in families of mechanical metamaterials where a duality transformation relates pairs of different configurations.
We show on general grounds how the existence of non-spatial symmetries further constrains the elastic tensor, reducing the number of independent moduli. 
In systems exhibiting a duality transformation, the resulting constraints on the number of moduli are particularly stringent at the self-dual point but persist even away from it, in a way reminiscent of critical phenomena. \end{abstract}

\maketitle

Classical elasticity describes how rigid objects respond to deformations~\cite{Hooke1678,LandauElasticity,Truesdell1960,Love1944,Chaikin1995}.
New facets of this time-honored subject continue to emerge in often unexpected guises and contexts. Recent examples range from quantum elasticity~\cite{Kleinert1989,Zaanen2004,Beekman2017} and fractons~\cite{Pretko2018,Gromov2019,Gromov2019b}, non-orientable elasticity~\cite{Bartolo2019}, the odd elasticity of active solids~\cite{Scheibner2019} and topological elasticity~\cite{Sun2012,Kane2013,Chen2014,Paulose2015,Lubensky2015,Huber2016,Rocklin2016,Po2016,Coulais2017,Sun2019,Saremi2019,Zhou2019,Rocklin2017}.

The very existence of rigid objects would seem rather mysterious if we were not so used to them in daily life: it is a consequence of the spontaneous breaking of translational invariance that occurs when a fluid condenses into a solid~\cite{Anderson1984}. This spontaneously broken symmetry guarantees the existence of excitations with arbitrarily low energies called Nambu-Goldstone modes~\cite{Nambu1960,Goldstone1961,Goldstone1962,Chaikin1995,Beekman2019}. In mechanics, the Goldstone modes are familiar objects: phonons of arbitrarily large wavelength~\cite{Leutwyler1996,Watanabe2020}. Elasticity can be viewed as the effective field theory of such Goldstone modes: a continuum description that ignores irrelevant microscopic details and instead focuses on the behavior at large scales relevant to our direct interactions with elastic bodies. 

The coarse-graining procedure that goes from a microscopic description to a continuum elastic theory should discard irrelevant details, but must crucially preserve symmetries~\cite{Curie1894,Goldenfeld1992}.
The spatial symmetries of a crystal can be gathered in a space group, containing all spatial transformations that leave the crystal invariant~\cite{BradleyCracknell,ITA}.
The space group of a crystal puts strong constraints on its elasticity, e.g., on the number of independent moduli~\cite{Nye1985,Teodosiu1982,LandauElasticity}. For instance, the elasticity of a two-dimensional crystal with triangular symmetry is isotropic (i.e., it is the same for all orientations) and, as a consequence, can display at most two independent elastic moduli.

In addition to spatial symmetries, additional non-spatial symmetries can occur microscopically. 
A symmetry is simply a transformation of the system that leaves it invariant. Symbolically, we can write $T(\mathcal{S}) = \mathcal{S}$ where~$\mathcal{S}$ represents the system, and~$T$ the symmetry transformation.
Spatial transformations such as rotations or translations can certainly be symmetries, but they do not exhaust all the possibilities.
Recent studies revealed that hidden non-spatial symmetries can emerge, for instance, in families of mechanical metamaterials where a duality transformation relates pairs of \emph{distinct} configurations~\cite{Fruchart2019}. Symbolically, two dual systems $\mathcal{S}_1$ and $\mathcal{S}_2$ related by the duality transformation $T$ satisfy $\mathcal{S}_2 = T(\mathcal{S}_1)$ and $\mathcal{S}_1 = T(\mathcal{S}_2)$.
The duality transformation has no reason to be a spatial transformation.
In self-dual systems (mapped onto themselves by the duality) the duality transformation can then become an additional hidden symmetry distinct from spatial ones.

In this Letter, we seek to determine the consequences of these additional constraints on the linear elasticity of a material.
More precisely, we consider the following question: how do microscopic symmetries affect the coarse-grained tensor of elastic moduli? 
Formally, we will determine the relation between the elastic tensors $c_{i j k \ell}$ and $\tilde{c}_{i j k \ell}$ of two systems $\mathcal{S}$ and $\tilde{\mathcal{S}} = T(\mathcal{S})$ respectively described by the momentum-space force-constant matrix~$S(q)$ and the transformed one $\tilde{S}(q) = U(q) S(u \cdot q) U(q)^{-1}$. Here, $S(q)$ relates the microscopic forces and displacements, while $U(q)$ and $u$ define the transformation (see next section for precise definitions).
For standard spatial symmetries, the answer is simply contained in the fact that $c_{i j k \ell}$ must transform as a tensor. 
Our analysis goes beyond this simple case and allows to analyze the effect of additional hidden (non-spatial) symmetries of the force-constant matrix, that can result in even stronger constraints. 
In addition, it applies to the case of dualities whereby the force-constant matrices of two different systems are related to each other by a nontrivial transformation. 

\begin{figure*}
  \centering
  \includegraphics{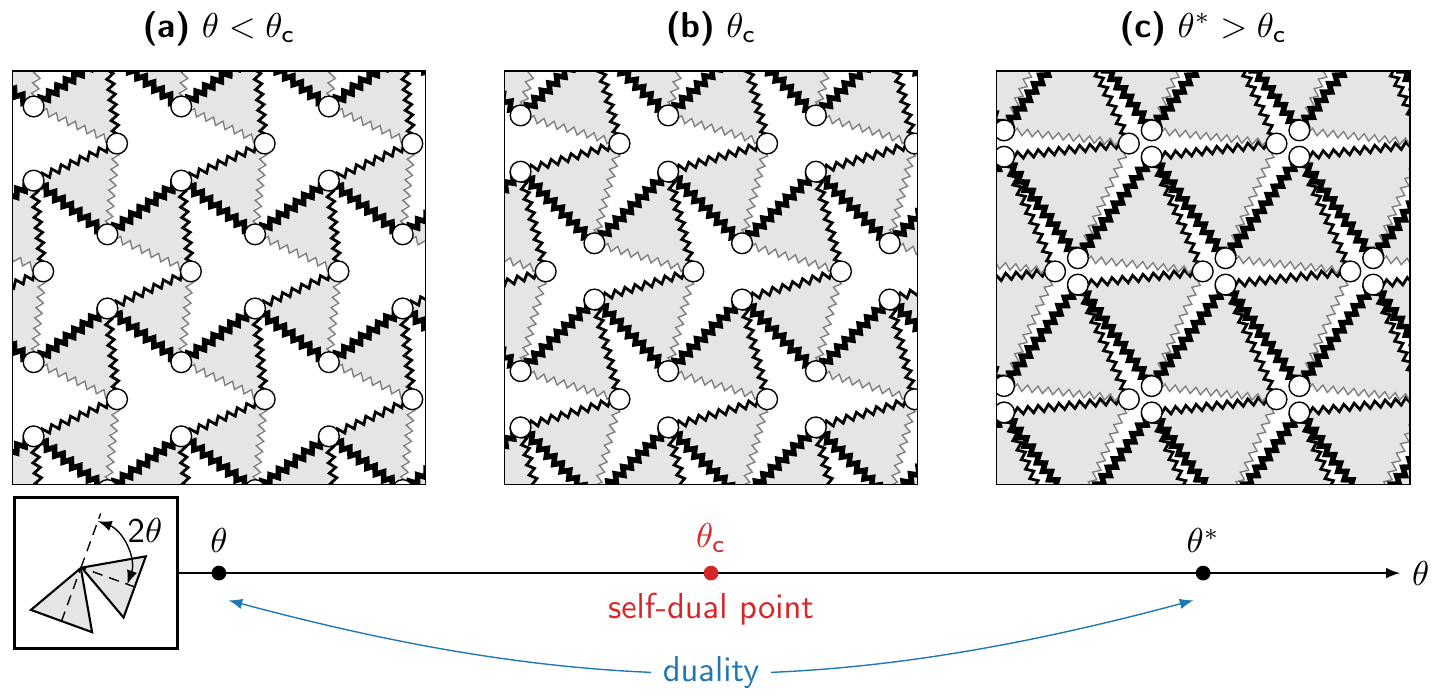}
  \caption{\label{elastic_kagome_lattice}\strong{Twisted Kagome lattice.}
  Examples of twisted Kagome lattices with different twisting angles.
  The duality maps structures with a twisting angle $\theta$ to ones with a twisting angle $\theta^* = 2 \theta_{\text{c}} - \theta$. 
  The critical structure with twisting angle $\theta_{\text{c}}$ is mapped to itself: it is self-dual.
  All these structures have the same space group, including the self-dual one.
  Here, inequivalent springs have different stiffnesses, as represented by their thicknesses in the figure, to remove any point group symmetry.
  (a) Below the critical angle.
  (b) At the critical angle.
  (c) Above the critical angle.
  Inset: definition of the twisting angle $\theta$.
  }
\end{figure*}

We apply our general formulas to the example of twisted Kagome lattices (see Fig.~\ref{elastic_kagome_lattice}), a family of two-dimensional crystals exhibiting a duality with a self-dual point where a non-spatial symmetry emerges~\cite{Fruchart2019}.
When all point group symmetries are lifted, six independent elastic moduli are expected in the continuum description of such systems.
Yet, the self-dual twisted Kagome lattices have isotropic elasticity with only one elastic modulus, despite not having any microscopic symmetry beyond Bravais lattice translations. Most strikingly, the elastic tensor is also constrained away from the self-dual point, reducing the number of elastic moduli to three. Our theory explains these counter-intuitive properties and casts them in a general formalism applicable beyond this concrete example.

\paragraphtitle{Linear Elasticity.} 
Linear elasticity describes the relation between the stress tensor $\sigma_{i j}$ and the displacement gradient (or strain) tensor $\epsilon_{k \ell}$, respectively representing the long-wavelength forces and deformations in a solid. 
More precisely, the displacement tensor is $\epsilon_{k \ell} = \partial u_{\ell} / \partial x_{k}$ where $u(x)$ represents the displacement of the point originally located at $x$ and now located at $X(x) = x + u(x)$, while the stress tensor is defined such that its divergence is the surface force $f_{i} = \partial_j \sigma_{i j}$ acting on an infinitesimal patch of material continuum.
We choose to work with the nonsymmetrized tensors to encompass recent extensions of elasticity where the antisymmetric components are relevant~\cite{Scheibner2019}.
Hooke's law in continuum form
\begin{equation}
	\label{hooke}
	\sigma_{i j} = c_{i j k \ell} \; \epsilon_{k \ell}
\end{equation}
linearly relates $\sigma_{i j}$ and $\epsilon_{k \ell}$ through the elastic tensor $c_{i j k \ell}$, whose entries are the static elastic moduli of the solid.

Spatial symmetries put strong constraints on the material properties of a crystal such as its elastic tensor $c_{i j k \ell}$.
This is because the elastic tensor $c_{i j k \ell}$ unsurprisingly transforms as a tensor under a spatial transformation $T \in O(d)$:
\begin{equation}
	\label{tensor_transformation_rule}
	c_{i j k \ell} \mapsto \tilde{c}_{i j k \ell} = T_{i i'} T_{j j'} T_{k k'} T_{\ell \ell'} c_{i' j' k' \ell'} .
\end{equation}
Hence, there is only a certain number of entries in $c_{i j k \ell}$ (i.e., of elastic moduli) that can be independent of each other, and those are prescribed by the symmetry of the material (we refer the reader to the SI and references therein for a short summary).
Yet, nothing guarantees that all of these moduli \emph{must} be independent, especially when additional constraints not originating from purely spatial symmetries exist.

Microscopically, we describe the elastic material as a set of massive particles arranged on a $d$-dimensional crystal and ruled by Newton equations $M \partial_t^2 u = F$,
where $u = x - x_{\text{eq}}$ are the displacements of the masses with respect to their equilibrium positions $x_{\text{eq}}$, and $M$ is a mass matrix describing the inertia of the particles. The forces $F$ between the particles are given in the harmonic approximation by $F = - S u$ where the force-constant matrix $S$ is essentially the matrix of second derivatives of the potential in the absence of pre-stress~\cite{Born1954}.
Hooke's law~\eqref{hooke} is the macroscopic version of the relation $F = - S u$ between forces and displacements.
Hence, the elastic tensor $c_{i j k \ell}$ can in principle be computed explicitly from the force-constant matrix $S$, see Ref.~\cite{Lutsko1989} (also Refs.~\cite{Lutsko1991,Barrat2006,Lemaitre2006,Zaccone2011,Liu2011}).

Here, we specialize to the case of a crystal, where particles are arranged in a spatially periodic fashion. 
Hence, we can use Bloch theorem to block-diagonalize Newton equations and to write $M \partial_t^2 u(q) = F(q) = - S(q) u(q)$ where $q$ is the quasi-momentum vector.
Because of the original translation invariance of the system (that is spontaneously broken), a global translation of the particles in any direction cannot induce any restoring force.
We assume that there is no other soft mode.
Hence, the kernel of the force-constant matrix $S(q=0)$ consists of the rigid-body translations of all the particles (i.e., the translations of the center of mass of the unit cell).
Elasticity describes the long-wavelength modes $q \to 0$ (acoustic phonons) projected onto rigid-body translations with the constraint that the projection of the force $F(q)$ on fast modes must relax (i.e., the projection on modes with a finite frequency at $q=0$, that span the orthogonal complement of the kernel, is zero).
The result of integrating out these irrelevant modes is in agreement with the the zero temperature limit of finite-temperature elasticity~\cite{Barrat2006}. 
The elastic tensor can then be obtained from the momentum-space force-constant matrix $S(q)$ near zero momentum as~\cite{Born1954,Lutsko1989,Scheibner2019}
\begin{equation}
	\label{elastic_tensor_from_force_constant_matrix}
 	\frac{c_{i j k \ell}}{\rho} = \left[ \frac{\partial^2 S}{\partial q_i \partial q_k} - \frac{\partial S}{\partial q_i} [S^{-1}] \frac{\partial S}{\partial q_k} \right]_{j \ell}
\end{equation}
where $\rho$ is the density, see \footnote{This expression is taken at momentum $q = 0$ and the inverse $S^{-1}$ is computed in the orthogonal complement of the kernel of the matrix}.
When all masses are equal, the force-constant matrix $S$ can be replaced by the more familiar dynamical matrix $D = M^{-1/2} \, S \, M^{-1/2}$. 

It is convenient to decompose the stress and deformation tensors in irreducible components. Hooke's law~\eqref{hooke} then reads~\cite{Avron1998,Scheibner2019} 
\begin{equation}
	\label{hooke_elastic_matrix}
	\sigma^{a} = K^{a b} \epsilon^b.
\end{equation}
In two dimensions, for instance, the four components of the stress (deformation) vector $\sigma^a$ ($\epsilon^{b}$) correspond to compression, rotation, and two linearly independent shear stresses (strains) [see SI for a visual representation].
More generally, $a$ and $b$ label basis matrices $\tau^{a}$ that span irreducible representations of $\text{SO}(d)$.
The elastic matrix $K^{a b} = {\footnotesize \frac{1}{4}} \sum_{i j k \ell} \tau^{a}_{i j} \, c_{i j k \ell} \, \tau^{b}_{k \ell}$ contains exactly the same information as the elastic tensor $c_{i j k \ell}$, only ordered in a different way.

\paragraphtitle{Symmetries and dualities and their effect on the elastic tensor.} 
We now consider a situation where a momentum-space force-constant matrix $\tilde{S}(q)$ is related to another force-constant matrix $S(q)$ by a relation of the form
\begin{equation}
	\label{duality_constraint}
	\tilde{S}(q) = (U S U^{-1})(\mathcal{O} \cdot q)
\end{equation}
where $U$ is unitary and $\mathcal{O}$ is orthogonal.
We stress that the matrices $U$ and $\mathcal{O}$ act on different spaces: $U$ acts on the displacements~$u$ of the masses, while $\mathcal{O}$ acts on the spatial coordinates $x$ (or equivalently momenta $q$).
This relation describes a symmetry when we impose $\tilde{S} = S$, i.e., the transformed system is identical to the original one.
It also describes situations where $\tilde{S}$ and $S$ are distinct, and in particular systems related by duality transformations ~\cite{Fruchart2019}.
The two force-constant matrices $S(q)$ and $\tilde{S}(q)$ define two elastic tensors $c_{i j k \ell}$ and $\tilde{c}_{i j k \ell}$ (equivalently, two elastic matrices $K^{a b}$ and $\tilde{K}^{a b}$) through equation~\eqref{elastic_tensor_from_force_constant_matrix}. 
We now proceed to determine the relation between $c_{i j k \ell}$ and $\tilde{c}_{i j k \ell}$ imposed by Eq.~\eqref{duality_constraint}.
Using equations \eqref{elastic_tensor_from_force_constant_matrix} and \eqref{duality_constraint}, one obtains by a direct calculation (see SI)
\begin{equation}
	\label{consequences_duality_elastic_tensor}
\tilde{c}_{i j k \ell} = \mathcal{O}_{i' i} \, R_{j j'} \, \mathcal{O}_{k' k} \, R_{\ell \ell'} \, c_{i' j' k' \ell'}
\end{equation}
where the orthogonal matrix $R$ is the projection of $U(0)$ on solid-body translations [the kernel of $S(0)$].
In terms of the elastic matrix $K$ in Eq.~\eqref{hooke_elastic_matrix}, the relation \eqref{consequences_duality_elastic_tensor} can be cast in the more compact form
\begin{equation}
	\tilde{K} = V K V^\dagger
\end{equation}
where
\begin{equation}
	\label{transformation_operator_elastic_matrix}
	V^{a b} = \frac{1}{2} \, \tr\left[ \tau^a \, R \, [\tau^b]^T \, \mathcal{O} \right].
\end{equation}

\medskip

\noindent The standard result Eq.~\eqref{tensor_transformation_rule} is recovered from Eq.~\eqref{consequences_duality_elastic_tensor} in the case of spatial symmetries, for which $R = \mathcal{O}^T \equiv T$. 
However, this particular case does not exhaust Eq.~\eqref{consequences_duality_elastic_tensor} as the relation \eqref{duality_constraint} is not necessarily the representation of a spatial symmetry, i.e. of an element of the space group of the crystal. As such, the matrix $R$ needs not be related to~$\mathcal{O}$ \footnote{
In Eq.~\eqref{consequences_duality_elastic_tensor}, half of the indices of $c_{i \mu j \nu}$ are transformed by $\mathcal{O}$ and the other half by $R$. 
In writing Eq.~\eqref{hooke}, we neglected the distinction between reference space (describing the undeformed elastic medium) and target space (describing the deformed medium), see Refs.~\cite{Truesdell1960,Truesdell2004,Ogden1997,Lubensky2002}. 
To emphasize this distinction, we use Latin indices for reference space coordinates $x_i$ and Greek indices for target space coordinates $X_\mu$.
The displacement gradient $\epsilon_{i \mu} = \partial u_\mu/\partial x_i$ and the stress $\sigma_{i \mu}$ are now objects with mixed indices (mixed/two-point tensors) and Hooke's law reads $\sigma_{i \mu} = c_{i \mu j \nu} \epsilon_{j \nu}$ in contrast with Eq.~\eqref{hooke}. 
This suggests that the elastic tensor $c_{i \mu j \nu}$ should not be restricted to transform according to Eq.~\eqref{tensor_transformation_rule}. Instead, different matrices can act on the reference and target spaces.
These considerations also lead to the less restrictive form in Eq.~\eqref{consequences_duality_elastic_tensor}. 
}.
In the next section, we shall present a concrete example where such hidden non-spatial symmetries occur in elasticity.

\paragraphtitle{Twisted Kagome lattices.}
Consider the family of mechanical structures called twisted Kagome lattices~\cite{Guest2003,Hutchinson2006,Sun2012,Kane2013,Lubensky2015,Paulose2015a}. 
These are two-dimensional periodic structures composed of three particles per unit cell on a triangular lattice, with each particle connected to four neighbors, as represented in Fig.~\ref{elastic_kagome_lattice}.
We consider a situation where inequivalent bonds (i.e., those not related by Bravais lattice translations) have different spring stiffnesses $k_i$, $i=1,2,3$ (see figure~\ref{elastic_kagome_lattice}).
This family is parametrized by a simple geometric parameter: the twisting angle~$\theta$ between two connected triangles, see the inset of Fig.~\ref{elastic_kagome_lattice}.
It was shown in Ref.~\cite{Fruchart2019} that a duality relates the dynamical matrices of the structures with $\theta$ and $\theta^* = 2 \theta_{\text{c}} - \theta$ (with $\theta_{\text{c}} = \pi/4$) through the relation~\cite{Fruchart2019}
\begin{equation}
  \label{unitary_nonlocal_duality}
  \mathscr{U}(k) D(\theta^*, -k) \mathscr{U}^{-1}(k) = D(\theta, k)
\end{equation}
where $\mathscr{U}(k) = \diag(\ii \varsigma_y, \ii \varsigma_y \ee^{- \ii k \cdot a_2}, \ii \varsigma_y \ee^{\ii k \cdot a_1})$.
In this expression, the matrices $\ii \varsigma_y$ act on the displacements $(x,y)$ of each of the three masses in the unit cell of the crystal, $\varsigma_i$ are Pauli matrices, and $a_i = [\cos((i-1) \, 2\pi/3), \sin((i-1) \, 2\pi/3)]^T$ are primitive vectors of the triangular Bravais lattice.
The duality \eqref{unitary_nonlocal_duality} typically relates different systems, with different twisting angles, such as the mechanical networks represented in Fig.~\ref{elastic_kagome_lattice} (a) and (c). However, there is a particular self-dual angle $\theta_{\text{c}} = \pi/4$ such that $\theta_{\text{c}}^* = \theta_{\text{c}}$ (see Fig.~\ref{elastic_kagome_lattice}), where the duality relation becomes an additional non-spatial symmetry of the dynamical matrix.

From Eq.~\eqref{unitary_nonlocal_duality}, one finds that $R = \ii \varsigma_y$ and $\mathcal{O} = - \Id$.
Upon substituting these results in Eq.~\eqref{transformation_operator_elastic_matrix}, we obtain 
\begin{equation}
	\label{V_matrix_Kagome}
  V = \varsigma_3 \otimes \ii \varsigma_2
\end{equation}
where $\otimes$ is the Kronecker product and $\varsigma_i$ are Pauli matrices. It is instructive to write the most general form of the elastic matrix for a standard material (i.e., energy and angular momentum are conserved and solid-body rotations do not change the elastic energy). 
In this situation, $K^{a 0} = 0 = K^{0 b}$ and $K^{a b} = K^{b a}$ (see Ref.~\cite{Scheibner2019} and SI for details), so we have
\begin{equation}
	\label{generic_standard_elastic_matrix}
	K = \begin{pmatrix}
		K^{0 0} & 0 & K^{0 2} & K^{0 3} \\
		0 		& 0 & 0       & 0       \\
		K^{0 2} & 0 & K^{2 2} & K^{2 3} \\
		K^{0 3} & 0 & K^{2 3} & K^{3 3} \\
	\end{pmatrix}.
\end{equation}

The elastic matrices $K(\theta)$ and $K(\theta^*)$ of two twisted Kagome lattices must indeed have the form~\eqref{generic_standard_elastic_matrix}. 
Following the preceding analysis, the duality relation \eqref{unitary_nonlocal_duality} implies an additional set of constraints
\begin{equation}
	\label{duality_elastic_moduli}
	V K(\theta) V^{\dagger} = K(\theta^*)
\end{equation}
with the transformation matrix $V$ defined in Eq.~\eqref{V_matrix_Kagome}. 
As a consequence, we find that
\begin{equation}
	\label{elasticity_matrix_kagome_theta}
	K(\theta) = \begin{pmatrix}
		0 & 0 & 0 & 0 \\
		0 		& 0 & 0       & 0       \\
		0 & 0 & K^{2 2}(\theta) & K^{2 3}(\theta) \\
		0 & 0 & K^{2 3}(\theta) & K^{3 3}(\theta) \\
	\end{pmatrix}
\end{equation}
with
\begin{subequations}
\begin{align}
	K^{22}(\theta) &= K^{33}(\theta^*) \\
	K^{33}(\theta) &= K^{22}(\theta^*) \\
	K^{23}(\theta) &= - K^{23}(\theta^*).
\end{align}
\end{subequations}
In particular, the constraint $V K(\theta_{\text{c}}) V^{\dagger} = K(\theta_{\text{c}})$ at the critical angle $\theta_{\text{c}} = \theta_{\text{c}}^*$ leads to $K^{22}(\theta_{\text{c}}) = K^{33}(\theta_{\text{c}})$ while $K^{23}(\theta_{\text{c}}) = 0$.

\begin{figure}
  \centering
  \includegraphics{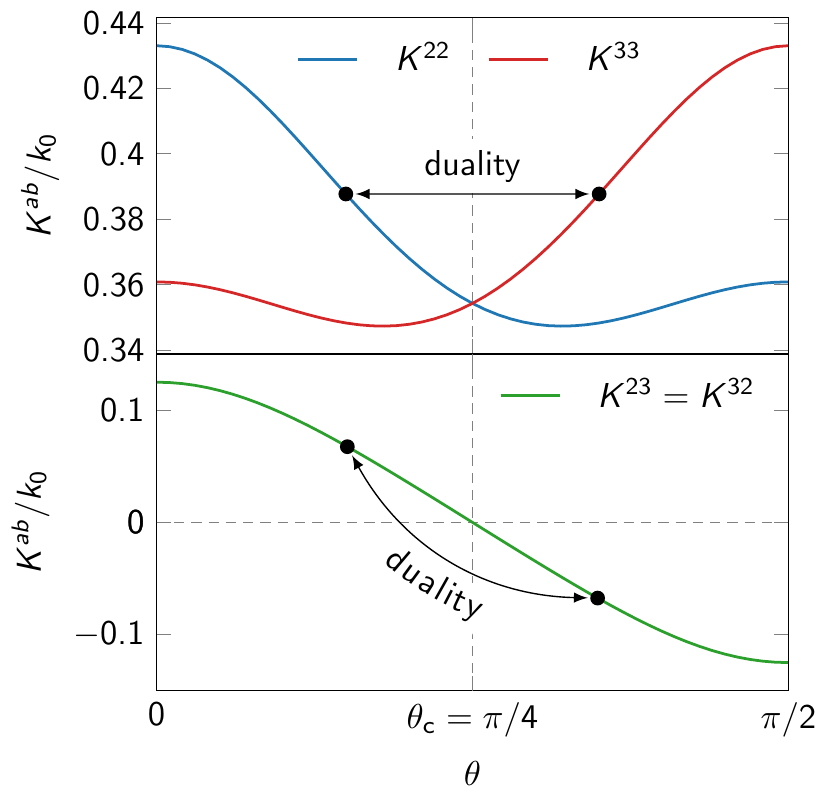}
  \caption{\label{figure_elastic_constants}\strong{Elastic constants for an anisotropic Kagome lattices.}
  The elastic moduli $K^{22}$, $K^{33}$ and $K^{23} = K^{32}$ computed from the microscopic description of Kagome lattices according to Eq.~\eqref{elastic_tensor_from_force_constant_matrix} are plotted as a function of the twisting angle $\theta$ for a generic situation where all inequivalent springs in the unit cell have different stiffnesses (see figure \ref{elastic_kagome_lattice})~\cite{Fruchart2019}.
  The duality (represented by black arrows) exchanges $K^{22}$ and $K^{33}$, as well as $K^{23}$ and $-K^{23}$.
  We have set $k_1 = k_0$, $k_1 = 2 k_0$, $k_3 = 3 k_0$.
  }
\end{figure}

Hence, the duality relation \eqref{duality_elastic_moduli} implies two striking consequences.
First, twisted Kagome lattices have only shear moduli: the coefficients $K^{0 0}$, $K^{0 2}$ and $K^{0 3}$ always vanish [see Eq.~\eqref{elasticity_matrix_kagome_theta}].
Crucially, the duality constrains the elastic moduli everywhere along the duality line (not only at the self-dual point).
Physically, the lack of bulk moduli is related to the existence of a Guest-Hutchinson mechanism~\cite{Guest2003,Hutchinson2006,Sun2012,Kane2013,Lubensky2015}, see in particular Ref.~\cite{Sun2012}.
Second, a stronger constraint occurs at the self-dual point where the elastic tensor becomes isotropic and characterized by a single shear modulus, despite no change in symmetry in the lattice. The occurrence of an isotropic elastic tensor holds even when all point group symmetries are lifted (i.e. the space group is \hm{p1}).
A direct computation of the elastic tensor from the dynamical matrix shown in Fig.~\ref{figure_elastic_constants}, using either Eq.~\eqref{elastic_tensor_from_force_constant_matrix} or the real-space equivalent~\cite{Lutsko1989} confirms all our results~\cite{Fruchart2019}.

\begin{figure}
  \centering
  \includegraphics{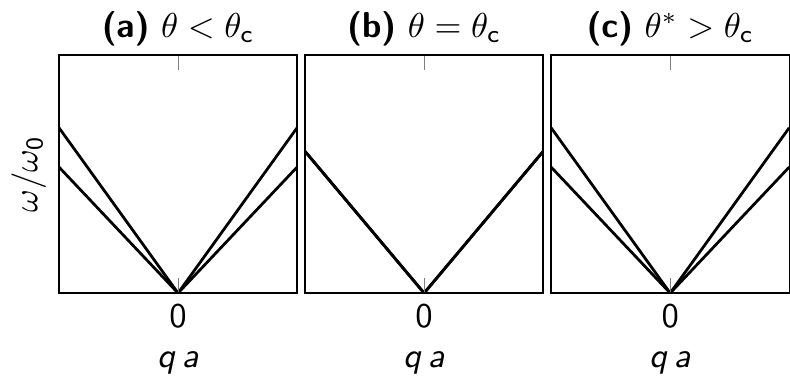}
  \caption{\label{figure_elastic_waves}\strong{Effect of dualities on elastic waves in an anisotropic Kagome lattices.}
  The dispersion relations of elastic waves in an anisotropic Kagome lattice are plotted for (a) $\theta=\num{0.1} \pi$, (b) $\theta_{\text{c}}$ and (c) the dual angle $\theta^*$ of case (a).
  The dispersion relations in (a) and (c) are identical, because of the duality between the corresponding systems.
  We distinguish the two acoustic branches in (a) or (c), but not in the self-dual system (b) where they share the same slope.
  We have set $k_1 = k_0$, $k_1 = 2 k_0$, $k_3 = 3 k_0$, $a$ is the lattice constant, and $\omega_{0}^2 = k_0/m$. The $x$ and $y$ axes have identical length.
  }
\end{figure}

To illustrate the effect of dualities, we consider the spectrum of elastic waves in anisotropic twisted Kagome lattices. 
The dynamics of elastic waves is described by the equation $\rho \, \ddot{u} = \nabla \cdot \sigma$. 
A Fourier transform of this equation gives $\omega^2 u_{i}(q) = h_{i \ell}(q) u_{\ell}(q)$ with $h_{i \ell}(q) = c_{i j k \ell} q_j q_k / \rho$. 
The dispersion relations obtained by diagonalizing the matrix $h(q)$ are plotted in Fig.~\ref{figure_elastic_waves}a-c.
We observe that the dual structures with twisting angles $\theta$ and $\theta^*$ (a and c) have identical spectra. 
Besides, the two branches are degenerate in the self-dual structure (b), as expected from the form of the elastic tensor.

\paragraphtitle{Conclusions.} 
We have shown how hidden non-spatial symmetries (originating, for instance, from dualities) strongly constrain the elastic moduli of a solid. 
Our results suggest a general mechanism not limited to elasticity by which microscopic dualities and non-spatial symmetries impose constraints on generalized rigidities and response functions.
These subtle effects are not captured by an analysis based on the spatial symmetry (i.e., the point group or space group) of the underlying structure. They are therefore likely to be overlooked in analyses performed purely within macroscopic continuum theories.

\appendix

\begin{center}
  \bf Appendices
\end{center}

\section{Number of elastic moduli}

Here, we recall a group-theoretic expression for the number of elastic moduli in a solid invariant under a certain point group.
In standard elasticity, the elastic tensor $c_{i j k \ell}$ is assumed to be invariant under the two so-called minor symmetries (i) $i \leftrightarrow j$ and (ii) $k \leftrightarrow \ell$, as well as under the so-called major symmetry (iii) $(i j) \leftrightarrow (k \ell)$ exchanging the unchanged groups of indices.
These correspond respectively to (i) the lack of antisymmetric stress (i.e. $\sigma_{i j} = \sigma_{j i}$) usually imposed by the conservation of angular momentum, (ii) the lack of coupling with rotations (i.e., solid-body rotations do not induce stresses), and (iii) the conservation of energy in the system. However, these hypotheses can be lifted in active systems (see Ref.~\cite{Scheibner2019} and Fig.~\ref{si_figure_elastic_matrix}). With such systems in mind, we also apply the general group-theoretic analysis to situations where the hypotheses (i)-(iii) do not hold, in the various cases summarized in Table~\ref{constraints_elastic_tensor}.

\newcommand{\cmark}{\ding{51}}\newcommand{\xmark}{\ding{55}}\newcommand{\oui}{\textcolor{tab_green}{\cmark}}\newcommand{\non}{\textcolor{tab_red}{\xmark}}

\begin{table*}
\centering 
\begin{tabular}{ccccccc}
\toprule
 & {$c_{i j k \ell}$} & {$c_{(i j) k \ell}$} & {$c_{i j (k \ell)}$} & {$c_{(i j) (k \ell)}$} & {$c_{(i j, k \ell)}$} & {$c_{((i j) (k \ell))}$} \\
\midrule 
energy conservation & \non & \non & \non & \non & \oui & \oui \\
angular momentum conservation & \non & \oui & \non & \oui & \non & \oui \\
no coupling with rotations & \non & \non & \oui & \oui & \non & \oui \\
\bottomrule
\end{tabular}
\caption{\label{constraints_elastic_tensor}\strong{Possible constraints on the elastic tensor.}
Indices or groups of indices in parentheses are symmetrized.
For instance, $2 c_{(i j) k \ell} =  c_{i j k \ell} + c_{j i k \ell} $, while $2 c_{(i j, k \ell)} = c_{i j k \ell} + c_{k \ell i j}$, etc.
}
\end{table*}

The number of independent elastic moduli is given by
\begin{equation}
	\label{number_of_elastic_moduli}
	n_1(\chi) = \frac{1}{|G|} \, \sum_{g \in G} \chi(g)
\end{equation}
where $G$ is the point group of the crystal, and $\chi$ is the character of the representation corresponding to the (potentially constrained) elastic tensor.
This quantity is computed for two-dimensional point groups in Table~\ref{number_elastic_moduli_2D} and for three-dimensional point groups in Table~\ref{number_elastic_moduli_3D} for the cases described in Table~\ref{constraints_elastic_tensor}.

The expression \enquote{the number of elastic moduli} is not entirely unambiguous: (a) we might decide to fix the axes of symmetry or not and (b) we might decide to rotate the crystal without changing the symmetry in order to reduce the number of nonzero coefficients.
The group-theoretical analysis assumes that the axes of symmetry are fixed (but do not necessarily coincide with the Cartesian axes used to express the elastic tensor), and gives the number of independent coefficients required to describe \emph{all} possible elastic tensors compatible with this data. 
Equivalently, this number assumes that the directions of the symmetry axes are independently known. 
In contrast, one could ask how many coefficients would be required to describe a crystal with a given point group, without knowing where the axes point to: the answer is in general different (this number is greater or equal to the one we compute).
One could also rotate a particular sample in order to reduce as much as possible the number of moduli, such as in Ref.~\cite[\S~10]{LandauElasticity} (this number is smaller or equal to the one we compute). 
Unlike the choice of a particular symmetry axis, that can in principle be done independently, this choice has nothing to do with symmetry: it is a choice of coordinate system along principal axes. 

\begin{table}
\centering 
\begin{tabular}{llS[table-column-width=1cm]S[table-column-width=1.1cm]S[table-column-width=1.1cm]S[table-column-width=1.2cm]S[table-column-width=1.2cm]} 
  \toprule
\multicolumn{2}{c}{point group} & {$c_{i j k \ell}$} & {$c_{(i j) k \ell}$} & {$c_{(i j) (k \ell)}$} & {$c_{(i j, k \ell)}$} & {$c_{((i j) (k \ell))}$} \\
  \midrule \hm{1} & \sch{C1} & 16 & 12 & 9 & 10 & 6 \\
\hm{2} & \sch{C2} & 16 & 12 & 9 & 10 & 6 \\
\hm{m} & \sch{Cs} & 8 & 6 & 5 & 6 & 4 \\
\hm{2mm} \phantom{gr} & \sch{C2v} & 8 & 6 & 5 & 6 & 4 \\
\hm{4} & \sch{C4} & 8 & 6 & 5 & 6 & 4 \\
\hm{4mm} & \sch{C4v} & 4 & 3 & 3 & 4 & 3 \\
\hm{3} & \sch{C3} & 6 & 4 & 3 & 4 & 2 \\
\hm{3m} & \sch{C3v} & 3 & 2 & 2 & 3 & 2 \\
\hm{6} & \sch{C6} & 6 & 4 & 3 & 4 & 2 \\
\hm{6mm} & \sch{C6v} & 3 & 2 & 2 & 3 & 2 \\
  \bottomrule \end{tabular}
\caption{\label{number_elastic_moduli_2D}\strong{Number of elastic moduli in 2D.}
The number of $c_{i j (k \ell)}$ is identical to the number of $c_{(i j) k \ell}$.
The number of moduli was computed with GAP~\cite{GAP} using the crystallographic database package CrystCat~\cite{CrystCat}.
The corresponding point groups are labeled with the conventions of Ref.~\cite{ITA} in Hermann-Mauguin notation (first column), as well as in Schoenflies notation (second column).
}
\end{table}

\begin{table}
\centering 
\begin{tabular}{llS[table-column-width=1cm]S[table-column-width=1.1cm]S[table-column-width=1.1cm]S[table-column-width=1.2cm]S[table-column-width=1.2cm]} 
  \toprule
\multicolumn{2}{c}{point group} & {$c_{i j k \ell}$} & {$c_{(i j) k \ell}$} & {$c_{(i j) (k \ell)}$} & {$c_{(i j, k \ell)}$} & {$c_{((i j) (k \ell))}$} \\
  \midrule \hm{1} & \sch{C1} & 81 & 54 & 36 & 45 & 21 \\
\hm{-1} & \sch{Ci} & 81 & 54 & 36 & 45 & 21 \\
\hm{2} & \sch{C2} & 41 & 28 & 20 & 25 & 13 \\
\hm{m} & \sch{Cs} & 41 & 28 & 20 & 25 & 13 \\
\hm{2/m} & \sch{C2h} & 41 & 28 & 20 & 25 & 13 \\
\hm{222} & \sch{D2} & 21 & 15 & 12 & 15 & 9 \\
\hm{mm2} & \sch{C2v} & 21 & 15 & 12 & 15 & 9 \\
\hm{mmm} & \sch{D2h} & 21 & 15 & 12 & 15 & 9 \\
\hm{4} & \sch{C4} & 21 & 14 & 10 & 13 & 7 \\
\hm{-4} & \sch{S4} & 21 & 14 & 10 & 13 & 7 \\
\hm{4/m} & \sch{C4h} & 21 & 14 & 10 & 13 & 7 \\
\hm{422} & \sch{D4} & 11 & 8 & 7 & 9 & 6 \\
\hm{4mm} & \sch{C4v} & 11 & 8 & 7 & 9 & 6 \\
\hm{-42m} & \sch{D2d} & 11 & 8 & 7 & 9 & 6 \\
\hm{4/mmm} & \sch{D4h} & 11 & 8 & 7 & 9 & 6 \\
\hm{3} & \sch{C3} & 27 & 18 & 12 & 15 & 7 \\
\hm{-3} & \sch{C3i} & 27 & 18 & 12 & 15 & 7 \\
\hm{32} & \sch{D3} & 14 & 10 & 8 & 10 & 6 \\
\hm{3m} & \sch{C3v} & 14 & 10 & 8 & 10 & 6 \\
\hm{-3m} & \sch{D3d} & 14 & 10 & 8 & 10 & 6 \\
\hm{6} & \sch{C6} & 19 & 12 & 8 & 11 & 5 \\
\hm{-6} & \sch{C3h} & 19 & 12 & 8 & 11 & 5 \\
\hm{6/m} & \sch{C6h} & 19 & 12 & 8 & 11 & 5 \\
\hm{622} & \sch{D6} & 10 & 7 & 6 & 8 & 5 \\
\hm{6mm} & \sch{C6v} & 10 & 7 & 6 & 8 & 5 \\
\hm{-62m} & \sch{D3h} & 10 & 7 & 6 & 8 & 5 \\
\hm{6/mmm} & \sch{D6h} & 10 & 7 & 6 & 8 & 5 \\
\hm{23} & \sch{T} & 7 & 5 & 4 & 5 & 3 \\
\hm{m-3} & \sch{Th} & 7 & 5 & 4 & 5 & 3 \\
\hm{432} & \sch{O} & 4 & 3 & 3 & 4 & 3 \\
\hm{-43m} & \sch{Td} & 4 & 3 & 3 & 4 & 3 \\
\hm{m-3m} & \sch{Oh} & 4 & 3 & 3 & 4 & 3 \\
  \bottomrule \end{tabular}
\caption{\label{number_elastic_moduli_3D}\strong{Number of elastic moduli in 3D.}
The number of $c_{i j (k \ell)}$ is identical to the number of $c_{(i j) k \ell}$.
The number of moduli was computed with GAP~\cite{GAP} using the crystallographic database package CrystCat~\cite{CrystCat}.
The corresponding point groups are labeled with the conventions of Ref.~\cite{ITA} in Hermann-Mauguin notation (first column), as well as in Schoenflies notation (second column).
}
\end{table}

\begin{table}
\centering 
\begin{tabular}{cl}
\toprule
tensor & $\chi(T)$ \\
\midrule 
{$c_{i j k \ell}$}      & $\displaystyle \tr(T)^4$  \\ \addlinespace[0.5em]
$\left.
\begin{tabular}{@{\ }l@{}}
    {$c_{(i j) k \ell}$} \\
    {$c_{i j (k \ell)}$}
 \end{tabular} \;
\right\}$ & $\displaystyle \frac{1}{2} \left[ \tr(T)^2 + \tr(T^2) \right] \tr(T)^2$  \\ \addlinespace[0.5em]
{$c_{(i j) (k \ell)}$}    & $\displaystyle \frac{1}{4} \left[ \tr(T)^2 + \tr(T^2) \right]^2$  \\ \addlinespace[0.5em]
{$c_{(i j, k \ell)}$}     & $\displaystyle \frac{1}{2} \left[ \tr(T)^4 + \tr(T^2)^2 \right]$  \\ \addlinespace[0.5em]
{$c_{((i j) (k \ell))}$}  & $\displaystyle \frac{1}{2} \left[  \frac{1}{4} \left( \tr(T)^2+\tr(T^2) \right)^2  + \frac{1}{2} \left(\tr(T^2)^2 + \tr(T^4) \right) \right]$  \\ \addlinespace[0.5em]
\bottomrule
\end{tabular}
\caption{\label{characters_representations_elastic_tensor}\strong{Characters of the different representations considered.}
}
\end{table}

\medskip

We now briefly describe how the standard formula \eqref{number_of_elastic_moduli} is obtained, and refer the reader to e.g. \cite[\S~VIII.41.2]{Lyubarskii1960} for more details (the same results can be obtain by direct inspection of the effects of group operations on the elastic tensor as explained e.g. in~\cite{Nye1985}; see also~\cite{Yang1994} for good summary of the method and its extension to quasicrystals).
The main idea in counting the number of independent coefficients is that the elastic tensor should be invariant under symmetries, and hence should transform according to the identity representation of the symmetry group (see e.g.~\cite{Tinkham1992,Dresselhaus2008,McWeeny1963,BradleyCracknell} for references on group theory).
First, we have to determine how the elastic tensor transforms under spatial transformations. This question can be rephrased as follows: under what representation $\Gamma$ of $O(d)$ (and hence of the point groups of interest) does the elastic tensor transform? 
We can then decompose $\Gamma$ into irreducible representations of the point group $G$; this decomposition looks like $\Gamma = n_1 \Gamma_1 \oplus n_2 \Gamma_2 \oplus \cdots \oplus n_N \Gamma_N$, where $\Gamma_i$ are the irreducible representations of $G$.
In general, such a decomposition means that there are $n_1$ basis tensors $T_1^{\alpha}$ (with $\alpha=1,\dots,n_1$) transforming under $\Gamma_1$, etc., such that
\begin{equation}
	c_{i j k \ell} = \sum_{k=1}^{N} \sum_{\alpha=1}^{n_k} c_{k}^{\alpha} [T_{k}^{\alpha}]_{i j k \ell}
\end{equation}
where $c_{k}^{\alpha}$ are the coefficients in the decomposition.
The elastic tensor must be invariant under the symmetry group $G$. 
This means that in this decomposition, only the part transforming along the identity representation $\Gamma_1$ (i.e., not transforming at all, as they are invariant) can stay.
Hence, there are $n_{1}$ independent elastic moduli.
To obtain this number explicitly, it is enough to know the character of the representation $\Gamma$, i.e. the trace of the representation applied to each element of the group.
The numbers $n_k(\chi)$ can be computed as
\begin{equation}
	n_k(\chi) = \frac{1}{|G|} \, \sum_{g \in G} \chi(g) \overline{\chi_k(g)}
\end{equation}
where $|G|$ is the number of elements in the group. 
The character of the identity representation is $\chi_1(g) = 1$ for all $g \in G$, which gives the equation \eqref{number_of_elastic_moduli}.

This number depends on the character $\chi$ of the representation $\Gamma$, which depends on our choices of constraints.
For instance, if we do not impose any symmetrization constraint on $c_{i j k \ell}$, then from the transformation rule
\begin{equation}
	\tilde{c}_{i j k \ell} = T_{i i'} T_{j j'} T_{k k'} T_{\ell \ell'} c_{i' j' k' \ell'}
\end{equation}
we infer that the character of the relevant representation (the rank four tensor representation) is 
\begin{equation}
	\chi(T) = T_{i i} T_{j j} T_{k k} T_{\ell \ell} = \tr(T)^4.
\end{equation}

However if we insist on having (for instance) $c_{i j k \ell} = c_{j i k \ell}$ then the linear transformation $T^{\otimes 4}$ must be similarly symmetrized, and we find
\begin{equation}
	\tilde{c}_{(i j) k \ell} = \frac{1}{2} \left[ T_{i i'} T_{j j'} + T_{i j'} T_{j i'} \right] T_{k k'} T_{\ell \ell'} c_{(i' j') k' \ell'}
\end{equation}
Hence, the trace (obtained by removing primes) gives the character
\begin{equation}
	\chi(T) = \frac{1}{2} \left[ \tr(T)^2 + \tr(T^2) \right] \tr(T)^2.
\end{equation}
The same procedure gives the characters of the representations corresponding to the cases described in Table~\ref{constraints_elastic_tensor}, and we summarize the results in Table~\ref{characters_representations_elastic_tensor}.

\section{Elastic matrix representation}

The elastic tensor can be decomposed as~\cite{Scheibner2019,Avron1998}
\begin{equation}
	c_{i j k \ell} = \sum_{a,b} K^{a b} \tau^{a}_{i j} \tau^{b}_{k \ell}.
\end{equation}
where the matrices $\tau^{\alpha} = \overline{\tau^{\alpha}}$ (seen as vectors) form a suitable orthonormal basis of $\mathcal{M}_{d}(\RR)$ (seen as a vector space endowed with a scalar product such as $\braket{M,N}=\tr(M^T N)/2$, as we will assume in the following).
Although any basis can formally be chosen, it is convenient to choose basis matrices from symmetry, see Ref.~\cite{Scheibner2019}.

The matrix $K^{a b}$ is obtained from the elastic tensor as
\begin{equation}
\frac{1}{4} \, \tau^{c}_{i j} \tau^{d}_{k \ell} c_{i j k \ell} 
	= \sum_{a,b} K^{a b}  \tau^{a}_{i j} \tau^{c}_{i j} \tau^{b}_{k \ell} \tau^{d}_{k \ell}
\end{equation}
where we recognize $\tau^{a}_{i j} \tau^{c}_{i j} = \tr([\tau^a]^T \tau^c) = 2 \braket{\tau^a, \tau^c} = 2 \delta^{a c}$ (similarly, we obtain $2 \delta^{b d}$). Hence,
\begin{equation}
	K^{a b} = \frac{1}{4} \, \sum_{i j k \ell} \tau^{a}_{i j} \tau^{b}_{k \ell} c_{i j k \ell}.
\end{equation}

\begin{figure*}
  \centering
  \includegraphics{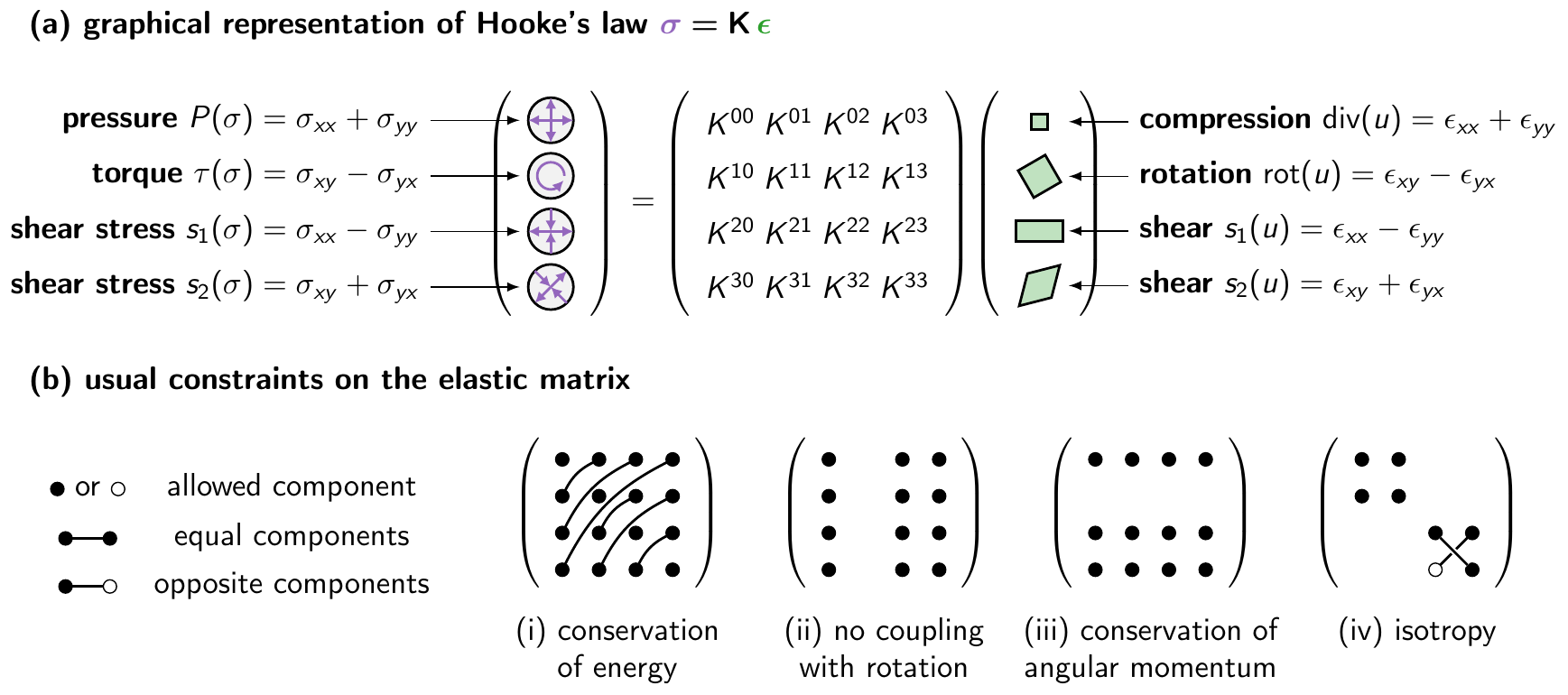}
  \caption{\label{si_figure_elastic_matrix}\strong{Semi-graphical representation of Hooke's law.}
  (a) Pictorial representation of Hooke's law~\cite{Scheibner2019}.
  (b) Some usual constraints in graphical form.
  (i) Conservation of energy implies $K^{a b} = K^{b a}$.
  (ii) Conservation of angular momentum implies $K^{1 b} = 0$.
  (iii) The lack of coupling with rotation (i.e. the fact that the state of stress of the material does not change after a solid body rotation) implies $K^{a 1} = 0$.
  (iv) Isotropy [defined here as the invariance under $\text{SO}(2)$] strongly reduces the number of elastic moduli from \num{16} to \num{6}. 
  Standard elasticity assumes that constraints (i-iii) are satisfied. 
  In isotropic standard elasticity where constraints (i-iv) are satisfied, there are only \num{2} elastic moduli, a shear modulus $K^{2 2} = K^{3 3}$ and a bulk modulus $K^{1 1}$.
  We refer to Ref.~\cite{Scheibner2019} for details and derivations.
  }
\end{figure*}

Figure~\ref{si_figure_elastic_matrix} gives a visual representation introduced in Ref.~\cite{Scheibner2019} of the elastic matrix in two dimensions, as well as visual representations of some usual constraints that can apply to the elastic matrix, also derived in Ref.~\cite{Scheibner2019}.

\section{Duality between elastic matrices in twisted Kagome lattices}

Consider the most general elastic matrix satisfying the constraints of standard elasticity (see figure~\ref{si_figure_elastic_matrix}) in two dimensions,
\begin{equation}
	\label{app_standard_material_K}
	K = \begin{pmatrix}
 K^{0 0} & 0 & K^{0 2} & K^{0 3} \\
 0 & 0 & 0 & 0 \\
 K^{0 2} & 0 & K^{2 2} & K^{2 3} \\
 K^{0 3} & 0 & K^{2 3} & K^{3 3} \\
	\end{pmatrix}.
\end{equation}
With the matrix
\begin{equation}
	\label{twisted_Kagome_V}
	V = \begin{pmatrix}
 0 & 1 & 0 & 0 \\
 -1 & 0 & 0 & 0 \\
 0 & 0 & 0 & -1 \\
 0 & 0 & 1 & 0 \\
	\end{pmatrix}
\end{equation}
we have
\begin{equation}
	V K V^\dagger = \begin{pmatrix}
 0 & 0 & 0 & 0 \\
 0 & K^{0 0} & K^{0 3} & -K^{0 2} \\
 0 & K^{0 3} & K^{3 3} & -K^{2 3} \\
 0 & -K^{0 2} & -K^{2 3} & K^{2 2} \\
	\end{pmatrix}.
\end{equation}

The duality relation (9) of the main text implies that for all $\theta$,
\begin{equation}
	V K(\theta) V^{\dagger} = K(\theta^*)
\end{equation}
where both $K(\theta)$ and $K(\theta^*)$ are constrained to be of the form \eqref{app_standard_material_K}.

Hence, we find that we always have
\begin{equation}
	K^{0 0}(\theta) = K^{0 2}(\theta) = K^{0 3}(\theta) = 0.
\end{equation}
Besides, the remaining coefficients at $\theta$ and $\theta^*$ are related through
\begin{subequations}
\begin{align}
	K^{22}(\theta) &= K^{33}(\theta^*) \\
	K^{33}(\theta) &= K^{22}(\theta^*) \\
	K^{23}(\theta) &= - K^{23}(\theta^*).
\end{align}
\end{subequations}
Hence, at the self-dual point, $K^{22}(\theta_{\text{c}}) = K^{33}(\theta_{\text{c}})$ and $K^{23}(\theta_{\text{c}}) = 0$.

\section{Determination of the elastic tensor from the microscopic description}
\def\idxi{i}
\def\idxj{j}

In this Appendix, we review the coarse-graining of the microscopic equations of motion summarized in the force constant matrix to the elastic tensor.
We refer the reader to Refs.~\cite{Scheibner2019,Lutsko1991,Lutsko1989,DiDonna2005,Barrat2006,Lemaitre2006,Maloney2006,Zaccone2011,Liu2011,Lubensky2015,Born1954} for more details.

We first diagonalize the momentum-space force constant matrix $S(0)$ at $q=0$. 
The corresponding basis of orthogonal eigenvectors are ordered by increasing eigenvalues, and $S(q)$ is written in the block form
\begin{equation}
    \label{blocks_force_constant_matrix}
	S(q) = \begin{pmatrix}
		S_{ZZ}(q) & S_{ZF}(q) \\
		S_{FZ}(q) & S_{FF}(q)
	\end{pmatrix}
\end{equation}
where the block $Z$ corresponds to the kernel of $S(0)$ i.e. to zero-frequency modes, while $F$ corresponds to the remaining modes with finite frequency.
By definition, $S(0)$ is block-diagonal and $S_{ZZ}(0) = 0$ while $S_{ZZ}(0) \equiv S_{ZZ}$ is invertible.

We further assume that $S_{ZZ}(q) = S_{ZZ}^{\idxi \idxj} q_{\idxi} q_{\idxj} + \mathcal{O}(q^2)$ i.e. that (i) there is no linear term in the series expansion of $S_{ZZ}(q)$ near $q=0$ and (ii) the second-order term is non-vanishing. 
This is not necessarily true: for instance, the system may be pre-stressed, or there may be lines of zero-frequency modes in the Brillouin zone (see e.g. Ref.~\cite{Lubensky2015}).
Degenerate cases can often be treated by adding a small perturbation to lift the degeneracy and taking the limit where this perturbation vanishes, provided that the limits for different perturbations are the same.

At lowest order in each block, the force-constant matrix then reads
\begin{equation}
    \label{lowest_order_force_constant_matrix}
	S(q) = \begin{pmatrix}
		S_{ZZ}^{\idxi \idxj} q_{\idxi} q_{\idxj} & S_{ZF}^{\idxi} q_{\idxi} \\
		S_{FZ}^{\idxj} q_{\idxj} & S_{FF}
	\end{pmatrix}
\end{equation}
where
\begin{equation}
	S_{ZZ}^{\idxi \idxj} = \left. \frac{\partial^2 S_{ZZ}}{\partial q_{\idxi} \, \partial q_{\idxj}} \right|_{q=0}
	\quad
	S_{ZF(FZ)}^{\idxi} = \left. \frac{\partial S_{ZF(FZ)}}{\partial q_{\idxi}} \right|_{q=0}
\end{equation}
and $S_{FF} = S_{FF}(q=0)$.

Let $u_{Z,i}$ with $i=1,\dots,d$ be a basis of the nullspace of $S(0)$.
These can usually be chosen as the solid-body motion of all particles in the unit cell in a given space direction (by construction), and labeled with spatial directions.
The displacement tensor (here written in momentum space) is related to the displacement field $u(q)$ by $\epsilon_{\idxi \idxj} = \ii \braket{u_{Z,\idxj}, \ii q_\idxi u(q)}$, i.e. it corresponds to the gradients of the projection on the nullspace of $S(0)$ of the displacements.

Similarly, the force (density) $f$ acting on the elastic body is identified to the projection $F_Z$ on the elastic degrees of freedom of the force $F = - S u$, so that $f_{\idxi}(q) = \rho F_{Z, \idxi}(q) = \ii q_\idxj \sigma_{\idxi \idxj}$.
We note that a general definition of the stress tensor requires some caution, especially when forces with long range (with respect to the microscopic scales in the lattice) are present (here, we avoided those issues by assuming a [possibly effective] description in terms of pairwise harmonic interactions).
First, a distinction has to be made between body forces and surface forces: we refer the reader to Refs.~\cite{Rinaldi2002,Israelachvili2011} for discussions.
Besides, the uniqueness of the stress tensor is a controversial issue (at first sight, it is uniquely defined only up to divergence free terms, but some additional assumptions appear to make it unique), and we refer the reader to Refs.~\cite{Irving1950,Schofield1982,Wajnryb1995,Goldhirsch2010} for more details.
We also refer to Ref.~\cite{Rao2020} for a similar discussion applied to viscosity coefficients in fluids.

The projection on the finite frequency part of the displacement is called non-affine displacement and is determined by assuming that the corresponding (non-elastic) projection of the force vanishes, $F_{F, \idxi} = 0$, see Refs. \cite{Lutsko1991,Lutsko1989,DiDonna2005,Barrat2006,Lemaitre2006,Maloney2006,Zaccone2011,Liu2011,Lubensky2015,Born1954}. 
In other words, we integrate out the irrelevant degrees of freedom at high frequency by solving for $u_F$ such that $F_{F, \idxi} = 0$ and replacing in the equations.
Physically, this is because the non-affine forces $F_{F, \idxi}$ relax due to thermal fluctuations: the additional term (called non-affine term) in the elastic tensor that accounts for this relaxation is the zero temperature limit of the term accounting for fluctuations in finite-temperature elasticity~\cite{Barrat2006}.

Hence, we have
\begin{equation}
	F = \begin{pmatrix}
		F_{Z} \\
		0 
	\end{pmatrix}
	= -
	\begin{pmatrix}
		q_{\idxi} & 0 \\
		0 & 1
	\end{pmatrix}
	\begin{pmatrix}
		S_{ZZ}^{\idxi \idxj} & S_{ZF}^{\idxi} \\
		S_{FZ}^{\idxj} & S_{FF}
	\end{pmatrix}
	\begin{pmatrix}
		q_{\idxj} & 0 \\
		0 & 1
	\end{pmatrix}
	\begin{pmatrix}
		u_Z \\
		u_F
	\end{pmatrix}
\end{equation}
where we have factorized the force-constant matrix. Hence,
\begin{equation}
	\begin{pmatrix}
		q_{\idxi}^{-1} F_{Z} \\
		0 
	\end{pmatrix}
	= -
	\begin{pmatrix}
		S_{ZZ}^{\idxi \idxj} & S_{ZF}^{\idxi} \\
		S_{FZ}^{\idxj} & S_{FF}
	\end{pmatrix}
	\begin{pmatrix}
		q_{\idxj} u_Z \\
		u_F
	\end{pmatrix}
\end{equation}
As multiplication by $q^{-1}$ corresponds to integration (effectively, we want to find $\sigma$ given $f$ and the relation $f = \text{Div}(\sigma)$), integration constants may in general appear~\cite{Scheibner2019}.
Here, we shall assume that such constants vanish.
Solving for $u_F$ and replacing then yields
\begin{equation}
	q_{\idxi}^{-1} F_{Z} = - [S_{ZZ}^{\idxi \idxj} - S_{ZF}^{\idxi} S_{FF}^{-1} S_{FZ}^{\idxj} ] q_{\idxj} u_Z
\end{equation}
where we recognize the deformation tensor and the stress tensor, so that
\begin{equation}
	\sigma_{\idxi m} = \rho [S_{ZZ}^{\idxi \idxj} - S_{ZF}^{\idxi} S_{FF}^{-1} S_{FZ}^{\idxj} ]_{m n} \epsilon_{\idxj n}.
\end{equation}
The elastic tensor is then
\begin{equation}
	\label{elastic_tensor_sigma}
	c_{i m j n} = \rho [\Sigma^{i j}]_{m n}
\end{equation}
where
\begin{equation}
	\Sigma^{i j} = S_{ZZ}^{i j} - S_{ZF}^{i} S_{FF}^{-1} S_{FZ}^{j}.
\end{equation}

\section{Symmetry of the elastic tensor for Hermitian force constant matrices}

When the momentum-space force constant matrix is Hermitian, the elastic tensor satisfies $c_{i j k \ell}=c_{k \ell i j}$ (equivalently, the elastic matrix is symmetric, $K = K^T$).
This constraint is related to energy conservation, see Ref.~\cite{Scheibner2019} and Table~\ref{constraints_elastic_tensor}, Figure~\ref{si_figure_elastic_matrix}.

This can be seen as follows: the momentum-space force constant matrix always satisfies $\overline{S}(q) = S(-q)$ (because the force constant matrix in physical space is real-valued). Assuming a Hermitian matrix [satisfying $S^\dagger(q) = S(q)$] implies that $S^T(q) = S(-q)$. 
As a consequence, the blocks in Eq.~\eqref{blocks_force_constant_matrix} satisfy
\begin{subequations}
\begin{align}
    S_{ZZ}^T(q) &= S_{ZZ}(-q) \\
    S_{FF}^T(q) &= S_{FF}(-q) \\
    S_{FZ}^T(q) &= S_{ZF}(-q).
\end{align}
\end{subequations}
Hence, the quantities defined in Eq.~\eqref{lowest_order_force_constant_matrix} satisfy
\begin{subequations}
\begin{align}
    [S_{ZZ}^{\idxi \idxj}]^T &= [S_{ZZ}^{\idxi \idxj}] \qquad
    &[S_{FF}]^T &= [S_{FF}] \\
    S_{FF}^T &= S_{FF} \qquad
    &[S_{FZ}^{\idxi}]^T &= - S_{ZF}^{\idxi}.
\end{align}
\end{subequations}
The symmetry of second derivatives gives $S_{ZZ}^{\idxi \idxj} = S_{ZZ}^{\idxj \idxi}$.
Putting all together, we obtain
\begin{equation}
	[\Sigma^{\idxi \idxj}]_{\alpha \beta} = [\Sigma^{\idxj \idxi}]_{\beta \alpha}
\end{equation}
which is (via Eq.~\eqref{elastic_tensor_sigma}) the announced symmetry of the elastic tensor.

\section{Symmetries of the elastic tensor from the symmetries of the force constant matrix}

We consider a force constant matrix $S(q)$, giving rise to an elastic tensor $c_{ijkl}$. 
Let us now consider the new force constant matrices (a) $\tilde{S}(q) = U(q) S(q) U(q)^{-1}$ and (b) $\tilde{S}(q) = S(u \cdot q)$ (we shall then combine the results) and determine the corresponding elastic tensors $\tilde{c}_{ijkl}$ in terms of the initial elastic tensor.

Consider first the case (a) where
\begin{equation}
	\tilde{S}(q) = U(q) S(q) U(q)^{-1}.
\end{equation}

For spatial symmetries, it is usually possible to assume that the symmetry operator does not depend on the momentum.
However, this is not the case for the duality operator considered in the main text. 
Hence, we must consider cases where the unitary matrix $U(q)$ depends explicitly on $q$.
However, we shall see that only the projection $U_{ZZ}$ to the kernel of $S(0)$ taken at $q=0$ appears in the transformation of the elastic tensor, as if only $U(0)$ was considered.
We first write the symmetry operator at the lowest non-trivial order in each block
\begin{equation}
	U(q) = \begin{pmatrix}
		U_{ZZ} & U_{ZF}^{\idxi} q_\idxi \\
		U_{FZ}^{\idxj} q_\idxj & U_{FF}
	\end{pmatrix}
\end{equation}
and
\begin{equation}
	U^{-1}(q) = \begin{pmatrix}
		U_{ZZ}^{-1} & \tilde{U}_{ZF}^{\idxi} q_\idxi \\
		\tilde{U}_{FZ}^{\idxj} q_\idxj & U_{FF}^{-1}.
	\end{pmatrix}
\end{equation}
At lowest order in each block,
\begin{equation}
	U(q) S(q) U^{-1}(q) =
	\begin{pmatrix}
		\tilde{S}_{ZZ}^{\idxi \idxj} q_{\idxi} q_{\idxj} & \tilde{S}_{ZF}^{\idxi} q_{\idxi} \\
		\tilde{S}_{FZ}^{\idxj} q_{\idxj} & \tilde{S}_{FF}
	\end{pmatrix}
\end{equation}
where
\begin{subequations}
\begin{align}
\begin{split}
	\tilde{S}_{ZZ}^{\idxi \idxj} &= 
	U_{ZZ} S^{\idxi \idxj} U_{ZZ}^{-1}
	 + U_{ZF}^{\idxi} S_{FZ}^{\idxj} U_{ZZ}^{-1} \\
	&\quad + U_{ZZ} S_{ZF}^{\idxi} \tilde{U}_{FZ}^{\idxj}
	 + U_{ZF}^{\idxi} S_{FF} \tilde{U}_{FZ}^{\idxj}
\end{split} \\
	\tilde{S}_{ZF}^{\idxi} &= U_{ZZ} S_{ZF}^{\idxi} U_{FF}^{-1} + U_{ZF}^{\idxi} S_{FF} U_{FF}^{-1} \\
	\tilde{S}_{FZ}^{\idxi} &= U_{FF} S_{FZ}^{\idxi} U_{ZZ}^{-1} + U_{FF} S_{FF} \tilde{U}_{FZ}^{\idxi} \\
	\tilde{S}_{FF} &= U_{FF} S_{FF} U_{FF}^{-1}
\end{align}
\end{subequations}
We also must have $\tilde{U}_{FZ}^{\idxi} = [U_{ZF}^{\idxi}]^{\dagger}$ so that $[\tilde{S}_{ZF}^{\idxi}]^\dagger = \tilde{S}_{FZ}^{\idxi}$.
Combining the preceding relations into
\begin{equation}
	\tilde{\Sigma}^{\idxi \idxj} = \tilde{S}_{ZZ}^{\idxi \idxj} - \tilde{S}_{ZF}^{\idxi} \tilde{S}_{FF}^{-1} \tilde{S}_{FZ}^{\idxj}
\end{equation}
yields, after simplification
\begin{equation}
	\tilde{\Sigma}^{\idxi \idxj} = U_{ZZ} \Sigma^{\idxi \idxj} U_{ZZ}^{-1}.
\end{equation}

Consider now the case (b) where
\begin{equation}
	\tilde{S}(q) = S(u \cdot q)
\end{equation}
where $u \in O(d)$ acts canonically on $q \in \RR^d$, namely so that $(u \cdot q)_{\idxi} = u_{\idxi \idxj} q_{\idxj}$.
A direct computation shows that the blocks indeed transform like tensors, namely
\begin{subequations}
\begin{align}
	\tilde{S}_{ZZ}^{\idxi \idxj} &= S_{ZZ}^{\idxi' \idxj'} u_{\idxi' \idxi} u_{\idxj' \idxj} \\
	\tilde{S}_{ZF}^{\idxi} &= S_{ZF}^{\idxi'} u_{\idxi' \idxi} \\
	\tilde{S}_{FZ}^{\idxj} &= S_{FZ}^{\idxj'} u_{\idxj' \idxj } \\
	\tilde{S}_{FF} &= S_{FF}
\end{align}
\end{subequations}
As a consequence,
\begin{equation}
	\tilde{\Sigma}^{\idxi \idxj} = \Sigma^{\rho \sigma} u_{\rho \idxi} u_{\sigma \idxj}
\end{equation}

Finally, consider the combination of cases (a) and (b),
\begin{equation}
	\tilde{S}(q) = (U S U^{-1})(u \cdot q).
\end{equation}
By combining the previous results, we obtain
\begin{equation}
	c_{i m j n} = [u^T]_{i i'} [U_{ZZ}]_{m m'} [u^T]_{j j'} [(U_{ZZ}^{-1})^T]_{n n'}  c_{i' m' j' n'}.
\end{equation}

Finally, let us write the constraint in terms of the elastic matrix. We want to compute
\begin{equation}
	\tilde{K}^{a b} = \frac{1}{4} \sum_{i j k l} \tau_{i j}^{a} \tilde{c}_{i j k l} \tau_{k l}^{b}
\end{equation}
in terms of
\begin{equation}
	K^{a b} = \frac{1}{4} \sum_{i j k l} \tau_{i j}^{a} c_{i j k l} \tau_{k l}^{b}.
\end{equation}

Using $c_{i' m' j' n'} = \tau^{c}_{i' m'} K^{c d} \tau^{d}_{j' n'}$ we get
\begin{equation}
	\tilde{K}^{a b} = V^{a c} K^{c d} W^{d b}
\end{equation}
with
\begin{align}
	V^{a c} &= \frac{1}{2} \tau^{a}_{i m} [U_{ZZ}]_{m m'} \tau^{c}_{i' m'} u_{i' i} \\
	W^{d b} &= \frac{1}{2} u_{j' j} \tau^{d}_{j' n'} [U_{ZZ}^{-1}]_{n' n} \tau^{b}_{j n}
\end{align}
Provided that $u \in O(d)$ and that $\tau$ matrices are real, $W^{d b} = \overline{V}^{b d}$, and we can write
\begin{equation}
	\tilde{K} = V K V^\dagger
\end{equation}
where (as defined above)
\begin{equation}
	V^{a c} = \frac{1}{2} \, \tr\left[ \tau^a U_{Z Z} [\tau^c]^T u \right].
\end{equation}
 

\vspace{0.75cm}


\begin{thebibliography}{72}\makeatletter
\providecommand \@ifxundefined [1]{\@ifx{#1\undefined}
}\providecommand \@ifnum [1]{\ifnum #1\expandafter \@firstoftwo
 \else \expandafter \@secondoftwo
 \fi
}\providecommand \@ifx [1]{\ifx #1\expandafter \@firstoftwo
 \else \expandafter \@secondoftwo
 \fi
}\providecommand \natexlab [1]{#1}\providecommand \enquote  [1]{``#1''}\providecommand \bibnamefont  [1]{#1}\providecommand \bibfnamefont [1]{#1}\providecommand \citenamefont [1]{#1}\providecommand \href@noop [0]{\@secondoftwo}\providecommand \href [0]{\begingroup \@sanitize@url \@href}\providecommand \@href[1]{\@@startlink{#1}\@@href}\providecommand \@@href[1]{\endgroup#1\@@endlink}\providecommand \@sanitize@url [0]{\catcode `\\12\catcode `\$12\catcode
  `\&12\catcode `\#12\catcode `\^12\catcode `\_12\catcode `\%12\relax}\providecommand \@@startlink[1]{}\providecommand \@@endlink[0]{}\providecommand \url  [0]{\begingroup\@sanitize@url \@url }\providecommand \@url [1]{\endgroup\@href {#1}{\urlprefix }}\providecommand \urlprefix  [0]{URL }\providecommand \Eprint [0]{\href }\providecommand \doibase [0]{https://doi.org/}\providecommand \selectlanguage [0]{\@gobble}\providecommand \bibinfo  [0]{\@secondoftwo}\providecommand \bibfield  [0]{\@secondoftwo}\providecommand \translation [1]{[#1]}\providecommand \BibitemOpen [0]{}\providecommand \bibitemStop [0]{}\providecommand \bibitemNoStop [0]{.\EOS\space}\providecommand \EOS [0]{\spacefactor3000\relax}\providecommand \BibitemShut  [1]{\csname bibitem#1\endcsname}\let\auto@bib@innerbib\@empty
\bibitem [{\citenamefont {Hooke}(1678)}]{Hooke1678}\BibitemOpen
  \bibfield  {author} {\bibinfo {author} {\bibfnamefont {R.}~\bibnamefont
  {Hooke}},\ }\href@noop {} {\emph {\bibinfo {title} {Lectures de Potentia
  Restitutiva, Or of Spring Explaining the Power of Springing Bodies}}}\
  (\bibinfo  {publisher} {John Martyn},\ \bibinfo {year} {1678})\BibitemShut
  {NoStop}\bibitem [{\citenamefont {Landau}\ and\ \citenamefont
  {Lifshitz}(1970)}]{LandauElasticity}\BibitemOpen
  \bibfield  {author} {\bibinfo {author} {\bibfnamefont {L.}~\bibnamefont
  {Landau}}\ and\ \bibinfo {author} {\bibfnamefont {E.}~\bibnamefont
  {Lifshitz}},\ }\href@noop {} {\emph {\bibinfo {title} {Theory of
  Elasticity}}},\ \bibinfo {number} {vol. 7}\ (\bibinfo  {publisher} {Pergamon
  Press},\ \bibinfo {year} {1970})\BibitemShut {NoStop}\bibitem [{\citenamefont {Truesdell}\ and\ \citenamefont
  {Toupin}(1960)}]{Truesdell1960}\BibitemOpen
  \bibfield  {author} {\bibinfo {author} {\bibfnamefont {C.}~\bibnamefont
  {Truesdell}}\ and\ \bibinfo {author} {\bibfnamefont {R.}~\bibnamefont
  {Toupin}},\ }\bibfield  {title} {\bibinfo {title} {{The Classical Field
  Theories}},\ }in\ \href {https://doi.org/10.1007/978-3-642-45943-6_2} {\emph
  {\bibinfo {booktitle} {Principles of Classical Mechanics and Field Theory /
  Prinzipien der Klassischen Mechanik und Feldtheorie}}}\ (\bibinfo
  {publisher} {Springer Berlin Heidelberg},\ \bibinfo {year} {1960})\ pp.\
  \bibinfo {pages} {226--858}\BibitemShut {NoStop}\bibitem [{\citenamefont {Love}(1944)}]{Love1944}\BibitemOpen
  \bibfield  {author} {\bibinfo {author} {\bibfnamefont {A.~E.~H.}\
  \bibnamefont {Love}},\ }\href@noop {} {\emph {\bibinfo {title} {A Treatise on
  the Mathematical Theory of Elasticity}}}\ (\bibinfo  {publisher} {Dover
  Publications},\ \bibinfo {year} {1944})\BibitemShut {NoStop}\bibitem [{\citenamefont {Chaikin}\ and\ \citenamefont
  {Lubensky}(1995)}]{Chaikin1995}\BibitemOpen
  \bibfield  {author} {\bibinfo {author} {\bibfnamefont {P.~M.}\ \bibnamefont
  {Chaikin}}\ and\ \bibinfo {author} {\bibfnamefont {T.~C.}\ \bibnamefont
  {Lubensky}},\ }\href {https://doi.org/10.1017/cbo9780511813467} {\emph
  {\bibinfo {title} {Principles of Condensed Matter Physics}}}\ (\bibinfo
  {publisher} {Cambridge University Press},\ \bibinfo {year}
  {1995})\BibitemShut {NoStop}\bibitem [{\citenamefont {Kleinert}(1989)}]{Kleinert1989}\BibitemOpen
  \bibfield  {author} {\bibinfo {author} {\bibfnamefont {H.}~\bibnamefont
  {Kleinert}},\ }\href {https://doi.org/10.1142/0356} {\emph {\bibinfo {title}
  {{Gauge Fields in Condensed Matter}}}}\ (\bibinfo  {publisher} {World
  Scientific},\ \bibinfo {year} {1989})\BibitemShut {NoStop}\bibitem [{\citenamefont {Zaanen}\ \emph {et~al.}(2004)\citenamefont {Zaanen},
  \citenamefont {Nussinov},\ and\ \citenamefont {Mukhin}}]{Zaanen2004}\BibitemOpen
  \bibfield  {author} {\bibinfo {author} {\bibfnamefont {J.}~\bibnamefont
  {Zaanen}}, \bibinfo {author} {\bibfnamefont {Z.}~\bibnamefont {Nussinov}},\
  and\ \bibinfo {author} {\bibfnamefont {S.}~\bibnamefont {Mukhin}},\
  }\bibfield  {title} {\bibinfo {title} {{Duality in 2+1D quantum elasticity:
  superconductivity and quantum nematic order}},\ }\href
  {https://doi.org/10.1016/j.aop.2003.10.003} {\bibfield  {journal} {\bibinfo
  {journal} {Annals of Physics}\ }\textbf {\bibinfo {volume} {310}},\ \bibinfo
  {pages} {181} (\bibinfo {year} {2004})}\BibitemShut {NoStop}\bibitem [{\citenamefont {Beekman}\ \emph {et~al.}(2017)\citenamefont
  {Beekman}, \citenamefont {Nissinen}, \citenamefont {Wu}, \citenamefont {Liu},
  \citenamefont {Slager}, \citenamefont {Nussinov}, \citenamefont {Cvetkovic},\
  and\ \citenamefont {Zaanen}}]{Beekman2017}\BibitemOpen
  \bibfield  {author} {\bibinfo {author} {\bibfnamefont {A.~J.}\ \bibnamefont
  {Beekman}}, \bibinfo {author} {\bibfnamefont {J.}~\bibnamefont {Nissinen}},
  \bibinfo {author} {\bibfnamefont {K.}~\bibnamefont {Wu}}, \bibinfo {author}
  {\bibfnamefont {K.}~\bibnamefont {Liu}}, \bibinfo {author} {\bibfnamefont
  {R.-J.}\ \bibnamefont {Slager}}, \bibinfo {author} {\bibfnamefont
  {Z.}~\bibnamefont {Nussinov}}, \bibinfo {author} {\bibfnamefont
  {V.}~\bibnamefont {Cvetkovic}},\ and\ \bibinfo {author} {\bibfnamefont
  {J.}~\bibnamefont {Zaanen}},\ }\bibfield  {title} {\bibinfo {title} {{Dual
  gauge field theory of quantum liquid crystals in two dimensions}},\ }\href
  {https://doi.org/10.1016/j.physrep.2017.03.004} {\bibfield  {journal}
  {\bibinfo  {journal} {Physics Reports}\ }\textbf {\bibinfo {volume} {683}},\
  \bibinfo {pages} {1} (\bibinfo {year} {2017})}\BibitemShut {NoStop}\bibitem [{\citenamefont {Pretko}\ and\ \citenamefont
  {Radzihovsky}(2018)}]{Pretko2018}\BibitemOpen
  \bibfield  {author} {\bibinfo {author} {\bibfnamefont {M.}~\bibnamefont
  {Pretko}}\ and\ \bibinfo {author} {\bibfnamefont {L.}~\bibnamefont
  {Radzihovsky}},\ }\bibfield  {title} {\bibinfo {title} {{Fracton-Elasticity
  Duality}},\ }\href {https://doi.org/10.1103/physrevlett.120.195301}
  {\bibfield  {journal} {\bibinfo  {journal} {Physical Review Letters}\
  }\textbf {\bibinfo {volume} {120}},\ \bibinfo {pages} {195301} (\bibinfo
  {year} {2018})}\BibitemShut {NoStop}\bibitem [{\citenamefont {Gromov}(2019)}]{Gromov2019}\BibitemOpen
  \bibfield  {author} {\bibinfo {author} {\bibfnamefont {A.}~\bibnamefont
  {Gromov}},\ }\bibfield  {title} {\bibinfo {title} {{Chiral Topological
  Elasticity and Fracton Order}},\ }\href
  {https://doi.org/10.1103/physrevlett.122.076403} {\bibfield  {journal}
  {\bibinfo  {journal} {Physical Review Letters}\ }\textbf {\bibinfo {volume}
  {122}},\ \bibinfo {pages} {076403} (\bibinfo {year} {2019})}\BibitemShut
  {NoStop}\bibitem [{\citenamefont {Gromov}\ and\ \citenamefont
  {Surowka}(2020)}]{Gromov2019b}\BibitemOpen
  \bibfield  {author} {\bibinfo {author} {\bibfnamefont {A.}~\bibnamefont
  {Gromov}}\ and\ \bibinfo {author} {\bibfnamefont {P.}~\bibnamefont
  {Surowka}},\ }\bibfield  {title} {\bibinfo {title} {{On duality between
  Cosserat elasticity and fractons}},\ }\href
  {https://doi.org/10.21468/scipostphys.8.4.065} {\bibfield  {journal}
  {\bibinfo  {journal} {{SciPost} Physics}\ }\textbf {\bibinfo {volume} {8}},\
  (\bibinfo {year} {2020})}\BibitemShut {NoStop}\bibitem [{\citenamefont {Bartolo}\ and\ \citenamefont
  {Carpentier}(2019)}]{Bartolo2019}\BibitemOpen
  \bibfield  {author} {\bibinfo {author} {\bibfnamefont {D.}~\bibnamefont
  {Bartolo}}\ and\ \bibinfo {author} {\bibfnamefont {D.}~\bibnamefont
  {Carpentier}},\ }\bibfield  {title} {\bibinfo {title} {Topological elasticity
  of nonorientable ribbons},\ }\href
  {https://doi.org/10.1103/physrevx.9.041058} {\bibfield  {journal} {\bibinfo
  {journal} {Physical Review X}\ }\textbf {\bibinfo {volume} {9}},\ \bibinfo
  {pages} {041058} (\bibinfo {year} {2019})}\BibitemShut {NoStop}\bibitem [{\citenamefont {Scheibner}\ \emph {et~al.}(2020)\citenamefont
  {Scheibner}, \citenamefont {Souslov}, \citenamefont {Banerjee}, \citenamefont
  {Sur{\'{o}}wka}, \citenamefont {Irvine},\ and\ \citenamefont
  {Vitelli}}]{Scheibner2019}\BibitemOpen
  \bibfield  {author} {\bibinfo {author} {\bibfnamefont {C.}~\bibnamefont
  {Scheibner}}, \bibinfo {author} {\bibfnamefont {A.}~\bibnamefont {Souslov}},
  \bibinfo {author} {\bibfnamefont {D.}~\bibnamefont {Banerjee}}, \bibinfo
  {author} {\bibfnamefont {P.}~\bibnamefont {Sur{\'{o}}wka}}, \bibinfo {author}
  {\bibfnamefont {W.~T.~M.}\ \bibnamefont {Irvine}},\ and\ \bibinfo {author}
  {\bibfnamefont {V.}~\bibnamefont {Vitelli}},\ }\bibfield  {title} {\bibinfo
  {title} {Odd elasticity},\ }\href {https://doi.org/10.1038/s41567-020-0795-y}
  {\bibfield  {journal} {\bibinfo  {journal} {Nature Physics}\ }\textbf
  {\bibinfo {volume} {16}},\ \bibinfo {pages} {475} (\bibinfo {year}
  {2020})}\BibitemShut {NoStop}\bibitem [{\citenamefont {Sun}\ \emph {et~al.}(2012)\citenamefont {Sun},
  \citenamefont {Souslov}, \citenamefont {Mao},\ and\ \citenamefont
  {Lubensky}}]{Sun2012}\BibitemOpen
  \bibfield  {author} {\bibinfo {author} {\bibfnamefont {K.}~\bibnamefont
  {Sun}}, \bibinfo {author} {\bibfnamefont {A.}~\bibnamefont {Souslov}},
  \bibinfo {author} {\bibfnamefont {X.}~\bibnamefont {Mao}},\ and\ \bibinfo
  {author} {\bibfnamefont {T.~C.}\ \bibnamefont {Lubensky}},\ }\bibfield
  {title} {\bibinfo {title} {{Surface phonons, elastic response, and conformal
  invariance in twisted kagome lattices}},\ }\href
  {https://doi.org/10.1073/pnas.1119941109} {\bibfield  {journal} {\bibinfo
  {journal} {Proceedings of the National Academy of Sciences}\ }\textbf
  {\bibinfo {volume} {109}},\ \bibinfo {pages} {12369} (\bibinfo {year}
  {2012})}\BibitemShut {NoStop}\bibitem [{\citenamefont {Kane}\ and\ \citenamefont
  {Lubensky}(2013)}]{Kane2013}\BibitemOpen
  \bibfield  {author} {\bibinfo {author} {\bibfnamefont {C.~L.}\ \bibnamefont
  {Kane}}\ and\ \bibinfo {author} {\bibfnamefont {T.~C.}\ \bibnamefont
  {Lubensky}},\ }\bibfield  {title} {\bibinfo {title} {{Topological boundary
  modes in isostatic lattices}},\ }\href {https://doi.org/10.1038/nphys2835}
  {\bibfield  {journal} {\bibinfo  {journal} {Nature Physics}\ }\textbf
  {\bibinfo {volume} {10}},\ \bibinfo {pages} {39} (\bibinfo {year}
  {2013})}\BibitemShut {NoStop}\bibitem [{\citenamefont {g.~Chen}\ \emph {et~al.}(2014)\citenamefont
  {g.~Chen}, \citenamefont {Upadhyaya},\ and\ \citenamefont
  {Vitelli}}]{Chen2014}\BibitemOpen
  \bibfield  {author} {\bibinfo {author} {\bibfnamefont {B.~G.}\ \bibnamefont
  {g.~Chen}}, \bibinfo {author} {\bibfnamefont {N.}~\bibnamefont {Upadhyaya}},\
  and\ \bibinfo {author} {\bibfnamefont {V.}~\bibnamefont {Vitelli}},\
  }\bibfield  {title} {\bibinfo {title} {{Nonlinear conduction via solitons in
  a topological mechanical insulator}},\ }\href
  {https://doi.org/10.1073/pnas.1405969111} {\bibfield  {journal} {\bibinfo
  {journal} {Proceedings of the National Academy of Sciences}\ }\textbf
  {\bibinfo {volume} {111}},\ \bibinfo {pages} {13004} (\bibinfo {year}
  {2014})}\BibitemShut {NoStop}\bibitem [{\citenamefont {Paulose}\ \emph
  {et~al.}(2015{\natexlab{a}})\citenamefont {Paulose}, \citenamefont
  {Meeussen},\ and\ \citenamefont {Vitelli}}]{Paulose2015}\BibitemOpen
  \bibfield  {author} {\bibinfo {author} {\bibfnamefont {J.}~\bibnamefont
  {Paulose}}, \bibinfo {author} {\bibfnamefont {A.~S.}\ \bibnamefont
  {Meeussen}},\ and\ \bibinfo {author} {\bibfnamefont {V.}~\bibnamefont
  {Vitelli}},\ }\bibfield  {title} {\bibinfo {title} {{Selective buckling via
  states of self-stress in topological metamaterials}},\ }\href
  {https://doi.org/10.1073/pnas.1502939112} {\bibfield  {journal} {\bibinfo
  {journal} {Proceedings of the National Academy of Sciences}\ }\textbf
  {\bibinfo {volume} {112}},\ \bibinfo {pages} {7639} (\bibinfo {year}
  {2015}{\natexlab{a}})}\BibitemShut {NoStop}\bibitem [{\citenamefont {Lubensky}\ \emph {et~al.}(2015)\citenamefont
  {Lubensky}, \citenamefont {Kane}, \citenamefont {Mao}, \citenamefont
  {Souslov},\ and\ \citenamefont {Sun}}]{Lubensky2015}\BibitemOpen
  \bibfield  {author} {\bibinfo {author} {\bibfnamefont {T.~C.}\ \bibnamefont
  {Lubensky}}, \bibinfo {author} {\bibfnamefont {C.~L.}\ \bibnamefont {Kane}},
  \bibinfo {author} {\bibfnamefont {X.}~\bibnamefont {Mao}}, \bibinfo {author}
  {\bibfnamefont {A.}~\bibnamefont {Souslov}},\ and\ \bibinfo {author}
  {\bibfnamefont {K.}~\bibnamefont {Sun}},\ }\bibfield  {title} {\bibinfo
  {title} {{Phonons and elasticity in critically coordinated lattices}},\
  }\href {https://doi.org/10.1088/0034-4885/78/7/073901} {\bibfield  {journal}
  {\bibinfo  {journal} {Reports on Progress in Physics}\ }\textbf {\bibinfo
  {volume} {78}},\ \bibinfo {pages} {073901} (\bibinfo {year}
  {2015})}\BibitemShut {NoStop}\bibitem [{\citenamefont {Huber}(2016)}]{Huber2016}\BibitemOpen
  \bibfield  {author} {\bibinfo {author} {\bibfnamefont {S.~D.}\ \bibnamefont
  {Huber}},\ }\bibfield  {title} {\bibinfo {title} {{Topological mechanics}},\
  }\href {https://doi.org/10.1038/nphys3801} {\bibfield  {journal} {\bibinfo
  {journal} {Nature Physics}\ }\textbf {\bibinfo {volume} {12}},\ \bibinfo
  {pages} {621} (\bibinfo {year} {2016})}\BibitemShut {NoStop}\bibitem [{\citenamefont {Rocklin}\ \emph {et~al.}(2016)\citenamefont
  {Rocklin}, \citenamefont {Chen}, \citenamefont {Falk}, \citenamefont
  {Vitelli},\ and\ \citenamefont {Lubensky}}]{Rocklin2016}\BibitemOpen
  \bibfield  {author} {\bibinfo {author} {\bibfnamefont {D.~Z.}\ \bibnamefont
  {Rocklin}}, \bibinfo {author} {\bibfnamefont {B.~G.}\ \bibnamefont {Chen}},
  \bibinfo {author} {\bibfnamefont {M.}~\bibnamefont {Falk}}, \bibinfo {author}
  {\bibfnamefont {V.}~\bibnamefont {Vitelli}},\ and\ \bibinfo {author}
  {\bibfnamefont {T.}~\bibnamefont {Lubensky}},\ }\bibfield  {title} {\bibinfo
  {title} {{Mechanical Weyl Modes in Topological Maxwell Lattices}},\ }\href
  {https://doi.org/10.1103/physrevlett.116.135503} {\bibfield  {journal}
  {\bibinfo  {journal} {Physical Review Letters}\ }\textbf {\bibinfo {volume}
  {116}},\ \bibinfo {pages} {135503} (\bibinfo {year} {2016})}\BibitemShut
  {NoStop}\bibitem [{\citenamefont {Po}\ \emph {et~al.}(2016)\citenamefont {Po},
  \citenamefont {Bahri},\ and\ \citenamefont {Vishwanath}}]{Po2016}\BibitemOpen
  \bibfield  {author} {\bibinfo {author} {\bibfnamefont {H.~C.}\ \bibnamefont
  {Po}}, \bibinfo {author} {\bibfnamefont {Y.}~\bibnamefont {Bahri}},\ and\
  \bibinfo {author} {\bibfnamefont {A.}~\bibnamefont {Vishwanath}},\ }\bibfield
   {title} {\bibinfo {title} {{Phonon analog of topological nodal
  semimetals}},\ }\href {https://doi.org/10.1103/physrevb.93.205158} {\bibfield
   {journal} {\bibinfo  {journal} {Physical Review B}\ }\textbf {\bibinfo
  {volume} {93}},\ \bibinfo {pages} {205158} (\bibinfo {year}
  {2016})}\BibitemShut {NoStop}\bibitem [{\citenamefont {Coulais}\ \emph {et~al.}(2017)\citenamefont
  {Coulais}, \citenamefont {Sounas},\ and\ \citenamefont
  {Al{\`{u}}}}]{Coulais2017}\BibitemOpen
  \bibfield  {author} {\bibinfo {author} {\bibfnamefont {C.}~\bibnamefont
  {Coulais}}, \bibinfo {author} {\bibfnamefont {D.}~\bibnamefont {Sounas}},\
  and\ \bibinfo {author} {\bibfnamefont {A.}~\bibnamefont {Al{\`{u}}}},\
  }\bibfield  {title} {\bibinfo {title} {{Static non-reciprocity in mechanical
  metamaterials}},\ }\href {https://doi.org/10.1038/nature21044} {\bibfield
  {journal} {\bibinfo  {journal} {Nature}\ }\textbf {\bibinfo {volume} {542}},\
  \bibinfo {pages} {461} (\bibinfo {year} {2017})}\BibitemShut {NoStop}\bibitem [{\citenamefont {Sun}\ and\ \citenamefont {Mao}(2019)}]{Sun2019}\BibitemOpen
  \bibfield  {author} {\bibinfo {author} {\bibfnamefont {K.}~\bibnamefont
  {Sun}}\ and\ \bibinfo {author} {\bibfnamefont {X.}~\bibnamefont {Mao}},\
  }\href@noop {} {\bibinfo {title} {{Universal Continuum Theory for Topological
  Edge Soft Modes}}} (\bibinfo {year} {2019}),\ \Eprint
  {https://arxiv.org/abs/1907.13163} {arXiv:1907.13163} \BibitemShut {NoStop}\bibitem [{\citenamefont {Saremi}\ and\ \citenamefont
  {Rocklin}(2020)}]{Saremi2019}\BibitemOpen
  \bibfield  {author} {\bibinfo {author} {\bibfnamefont {A.}~\bibnamefont
  {Saremi}}\ and\ \bibinfo {author} {\bibfnamefont {Z.}~\bibnamefont
  {Rocklin}},\ }\bibfield  {title} {\bibinfo {title} {Topological elasticity of
  flexible structures},\ }\href {https://doi.org/10.1103/physrevx.10.011052}
  {\bibfield  {journal} {\bibinfo  {journal} {Physical Review X}\ }\textbf
  {\bibinfo {volume} {10}},\ \bibinfo {pages} {011052} (\bibinfo {year}
  {2020})}\BibitemShut {NoStop}\bibitem [{\citenamefont {Zhou}\ \emph {et~al.}(2019)\citenamefont {Zhou},
  \citenamefont {Zhang},\ and\ \citenamefont {Mao}}]{Zhou2019}\BibitemOpen
  \bibfield  {author} {\bibinfo {author} {\bibfnamefont {D.}~\bibnamefont
  {Zhou}}, \bibinfo {author} {\bibfnamefont {L.}~\bibnamefont {Zhang}},\ and\
  \bibinfo {author} {\bibfnamefont {X.}~\bibnamefont {Mao}},\ }\bibfield
  {title} {\bibinfo {title} {Topological boundary floppy modes in
  quasicrystals},\ }\href {https://doi.org/10.1103/physrevx.9.021054}
  {\bibfield  {journal} {\bibinfo  {journal} {Physical Review X}\ }\textbf
  {\bibinfo {volume} {9}},\ \bibinfo {pages} {021054} (\bibinfo {year}
  {2019})}\BibitemShut {NoStop}\bibitem [{\citenamefont {Rocklin}\ \emph {et~al.}(2017)\citenamefont
  {Rocklin}, \citenamefont {Zhou}, \citenamefont {Sun},\ and\ \citenamefont
  {Mao}}]{Rocklin2017}\BibitemOpen
  \bibfield  {author} {\bibinfo {author} {\bibfnamefont {D.~Z.}\ \bibnamefont
  {Rocklin}}, \bibinfo {author} {\bibfnamefont {S.}~\bibnamefont {Zhou}},
  \bibinfo {author} {\bibfnamefont {K.}~\bibnamefont {Sun}},\ and\ \bibinfo
  {author} {\bibfnamefont {X.}~\bibnamefont {Mao}},\ }\bibfield  {title}
  {\bibinfo {title} {{Transformable topological mechanical metamaterials}},\
  }\href {https://doi.org/10.1038/ncomms14201} {\bibfield  {journal} {\bibinfo
  {journal} {Nature Communications}\ }\textbf {\bibinfo {volume} {8}},\
  \bibinfo {pages} {14201} (\bibinfo {year} {2017})}\BibitemShut {NoStop}\bibitem [{\citenamefont {Anderson}(1984)}]{Anderson1984}\BibitemOpen
  \bibfield  {author} {\bibinfo {author} {\bibfnamefont {P.~W.}\ \bibnamefont
  {Anderson}},\ }\href@noop {} {\emph {\bibinfo {title} {Basic notions of
  condensed matter physics}}}\ (\bibinfo  {publisher} {Westview Press},\
  \bibinfo {year} {1984})\BibitemShut {NoStop}\bibitem [{\citenamefont {Nambu}(1960)}]{Nambu1960}\BibitemOpen
  \bibfield  {author} {\bibinfo {author} {\bibfnamefont {Y.}~\bibnamefont
  {Nambu}},\ }\bibfield  {title} {\bibinfo {title} {{Quasi-Particles and Gauge
  Invariance in the Theory of Superconductivity}},\ }\href
  {https://doi.org/10.1103/physrev.117.648} {\bibfield  {journal} {\bibinfo
  {journal} {Physical Review}\ }\textbf {\bibinfo {volume} {117}},\ \bibinfo
  {pages} {648} (\bibinfo {year} {1960})}\BibitemShut {NoStop}\bibitem [{\citenamefont {Goldstone}(1961)}]{Goldstone1961}\BibitemOpen
  \bibfield  {author} {\bibinfo {author} {\bibfnamefont {J.}~\bibnamefont
  {Goldstone}},\ }\bibfield  {title} {\bibinfo {title} {{Field theories with
  Superconductor solutions}},\ }\href {https://doi.org/10.1007/bf02812722}
  {\bibfield  {journal} {\bibinfo  {journal} {Il Nuovo Cimento}\ }\textbf
  {\bibinfo {volume} {19}},\ \bibinfo {pages} {154} (\bibinfo {year}
  {1961})}\BibitemShut {NoStop}\bibitem [{\citenamefont {Goldstone}\ \emph {et~al.}(1962)\citenamefont
  {Goldstone}, \citenamefont {Salam},\ and\ \citenamefont
  {Weinberg}}]{Goldstone1962}\BibitemOpen
  \bibfield  {author} {\bibinfo {author} {\bibfnamefont {J.}~\bibnamefont
  {Goldstone}}, \bibinfo {author} {\bibfnamefont {A.}~\bibnamefont {Salam}},\
  and\ \bibinfo {author} {\bibfnamefont {S.}~\bibnamefont {Weinberg}},\
  }\bibfield  {title} {\bibinfo {title} {{Broken Symmetries}},\ }\href
  {https://doi.org/10.1103/physrev.127.965} {\bibfield  {journal} {\bibinfo
  {journal} {Physical Review}\ }\textbf {\bibinfo {volume} {127}},\ \bibinfo
  {pages} {965} (\bibinfo {year} {1962})}\BibitemShut {NoStop}\bibitem [{\citenamefont {Beekman}\ \emph {et~al.}(2019)\citenamefont
  {Beekman}, \citenamefont {Rademaker},\ and\ \citenamefont {van
  Wezel}}]{Beekman2019}\BibitemOpen
  \bibfield  {author} {\bibinfo {author} {\bibfnamefont {A.}~\bibnamefont
  {Beekman}}, \bibinfo {author} {\bibfnamefont {L.}~\bibnamefont {Rademaker}},\
  and\ \bibinfo {author} {\bibfnamefont {J.}~\bibnamefont {van Wezel}},\
  }\bibfield  {title} {\bibinfo {title} {{An introduction to spontaneous
  symmetry breaking}},\ }\href
  {https://doi.org/10.21468/scipostphyslectnotes.11} {\bibfield  {journal}
  {\bibinfo  {journal} {{SciPost} Physics Lecture Notes}\ }\textbf {\bibinfo
  {volume} {11}},\  (\bibinfo {year} {2019})}\BibitemShut {NoStop}\bibitem [{\citenamefont {Leutwyler}(1997)}]{Leutwyler1996}\BibitemOpen
  \bibfield  {author} {\bibinfo {author} {\bibfnamefont {H.}~\bibnamefont
  {Leutwyler}},\ }\bibfield  {title} {\bibinfo {title} {{Phonons as Goldstone
  Bosons}},\ }\href {https://doi.org/10.5169/seals-117020} {\bibfield
  {journal} {\bibinfo  {journal} {Helv. Phys. Acta}\ }\textbf {\bibinfo
  {volume} {70}},\ \bibinfo {pages} {275} (\bibinfo {year} {1997})},\ \Eprint
  {https://arxiv.org/abs/hep-ph/9609466} {arXiv:hep-ph/9609466} \BibitemShut
  {NoStop}\bibitem [{\citenamefont {Watanabe}(2020)}]{Watanabe2020}\BibitemOpen
  \bibfield  {author} {\bibinfo {author} {\bibfnamefont {H.}~\bibnamefont
  {Watanabe}},\ }\bibfield  {title} {\bibinfo {title} {{Counting Rules of
  Nambu-Goldstone Modes}},\ }\href
  {https://doi.org/10.1146/annurev-conmatphys-031119-050644} {\bibfield
  {journal} {\bibinfo  {journal} {Annual Review of Condensed Matter Physics}\
  }\textbf {\bibinfo {volume} {11}},\ \bibinfo {pages} {169} (\bibinfo {year}
  {2020})}\BibitemShut {NoStop}\bibitem [{\citenamefont {Curie}(1894)}]{Curie1894}\BibitemOpen
  \bibfield  {author} {\bibinfo {author} {\bibfnamefont {P.}~\bibnamefont
  {Curie}},\ }\bibfield  {title} {\bibinfo {title} {{{Sur la sym{\'e}trie dans
  les ph{\'e}nom{\`e}nes physiques, sym{\'e}trie d'un champ {\'e}lectrique et
  d'un champ magn{\'e}tique}}},\ }\href
  {https://doi.org/10.1051/jphystap:018940030039300} {\bibfield  {journal}
  {\bibinfo  {journal} {{J. Phys. Theor. Appl.}}\ }\textbf {\bibinfo {volume}
  {3}},\ \bibinfo {pages} {393} (\bibinfo {year} {1894})}\BibitemShut {NoStop}\bibitem [{\citenamefont {Goldenfeld}(1992)}]{Goldenfeld1992}\BibitemOpen
  \bibfield  {author} {\bibinfo {author} {\bibfnamefont {N.}~\bibnamefont
  {Goldenfeld}},\ }\href@noop {} {\emph {\bibinfo {title} {{Lectures on Phase
  Transitions and the Renormalization Group}}}}\ (\bibinfo  {publisher}
  {Perseus Books},\ \bibinfo {year} {1992})\BibitemShut {NoStop}\bibitem [{\citenamefont {Bradley}\ and\ \citenamefont
  {Cracknell}(2010)}]{BradleyCracknell}\BibitemOpen
  \bibfield  {author} {\bibinfo {author} {\bibfnamefont {C.}~\bibnamefont
  {Bradley}}\ and\ \bibinfo {author} {\bibfnamefont {A.}~\bibnamefont
  {Cracknell}},\ }\href@noop {} {\emph {\bibinfo {title} {The Mathematical
  Theory of Symmetry in Solids: Representation Theory for Point Groups and
  Space Groups}}}\ (\bibinfo  {publisher} {Oxford University Press},\ \bibinfo
  {year} {2010})\BibitemShut {NoStop}\bibitem [{\citenamefont {Aroyo}(2016)}]{ITA}\BibitemOpen
  \bibinfo {editor} {\bibfnamefont {M.~I.}\ \bibnamefont {Aroyo}},\ ed.,\ \href
  {https://doi.org/10.1107/97809553602060000114} {\emph {\bibinfo {title}
  {{International Tables for Crystallography, Volume A: Space-group
  symmetry}}}}\ (\bibinfo  {publisher} {International Union of
  Crystallography},\ \bibinfo {year} {2016})\BibitemShut {NoStop}\bibitem [{\citenamefont {Nye}(1985)}]{Nye1985}\BibitemOpen
  \bibfield  {author} {\bibinfo {author} {\bibfnamefont {J.~F.}\ \bibnamefont
  {Nye}},\ }\href@noop {} {\emph {\bibinfo {title} {Physical Properties Of
  Crystals: Their Representation by Tensors and Matrices}}}\ (\bibinfo
  {publisher} {Oxford University Press},\ \bibinfo {year} {1985})\BibitemShut
  {NoStop}\bibitem [{\citenamefont {Teodosiu}(1982)}]{Teodosiu1982}\BibitemOpen
  \bibfield  {author} {\bibinfo {author} {\bibfnamefont {C.}~\bibnamefont
  {Teodosiu}},\ }\href@noop {} {\emph {\bibinfo {title} {Elastic models of
  crystal defects}}}\ (\bibinfo  {publisher} {Springer},\ \bibinfo {year}
  {1982})\BibitemShut {NoStop}\bibitem [{\citenamefont {Fruchart}\ \emph {et~al.}(2020)\citenamefont
  {Fruchart}, \citenamefont {Zhou},\ and\ \citenamefont
  {Vitelli}}]{Fruchart2019}\BibitemOpen
  \bibfield  {author} {\bibinfo {author} {\bibfnamefont {M.}~\bibnamefont
  {Fruchart}}, \bibinfo {author} {\bibfnamefont {Y.}~\bibnamefont {Zhou}},\
  and\ \bibinfo {author} {\bibfnamefont {V.}~\bibnamefont {Vitelli}},\
  }\bibfield  {title} {\bibinfo {title} {Dualities and non-abelian mechanics},\
  }\href {https://doi.org/10.1038/s41586-020-1932-6} {\bibfield  {journal}
  {\bibinfo  {journal} {Nature}\ }\textbf {\bibinfo {volume} {577}},\ \bibinfo
  {pages} {636} (\bibinfo {year} {2020})}\BibitemShut {NoStop}\bibitem [{\citenamefont {Born}\ and\ \citenamefont {Huang}(1954)}]{Born1954}\BibitemOpen
  \bibfield  {author} {\bibinfo {author} {\bibfnamefont {M.}~\bibnamefont
  {Born}}\ and\ \bibinfo {author} {\bibfnamefont {K.}~\bibnamefont {Huang}},\
  }\href@noop {} {\emph {\bibinfo {title} {Dynamical theory of crystal
  lattices}}}\ (\bibinfo  {publisher} {Clarendon press},\ \bibinfo {year}
  {1954})\BibitemShut {NoStop}\bibitem [{\citenamefont {Lutsko}(1989)}]{Lutsko1989}\BibitemOpen
  \bibfield  {author} {\bibinfo {author} {\bibfnamefont {J.~F.}\ \bibnamefont
  {Lutsko}},\ }\bibfield  {title} {\bibinfo {title} {{Generalized expressions
  for the calculation of elastic constants by computer simulation}},\ }\href
  {https://doi.org/10.1063/1.342716} {\bibfield  {journal} {\bibinfo  {journal}
  {Journal of Applied Physics}\ }\textbf {\bibinfo {volume} {65}},\ \bibinfo
  {pages} {2991} (\bibinfo {year} {1989})}\BibitemShut {NoStop}\bibitem [{\citenamefont {Lutsko}(1991)}]{Lutsko1991}\BibitemOpen
  \bibfield  {author} {\bibinfo {author} {\bibfnamefont {J.~F.}\ \bibnamefont
  {Lutsko}},\ }\bibfield  {title} {\bibinfo {title} {{The Determination of the
  Elastic Properties of Inhomogeneous Systems by Computer Simulation}},\ }in\
  \href {https://doi.org/10.1007/978-94-011-3546-7_16} {\emph {\bibinfo
  {booktitle} {Computer Simulation in Materials Science}}}\ (\bibinfo
  {publisher} {Springer Netherlands},\ \bibinfo {year} {1991})\ pp.\ \bibinfo
  {pages} {335--348}\BibitemShut {NoStop}\bibitem [{\citenamefont {Barrat}(2006)}]{Barrat2006}\BibitemOpen
  \bibfield  {author} {\bibinfo {author} {\bibfnamefont {J.-L.}\ \bibnamefont
  {Barrat}},\ }\bibfield  {title} {\bibinfo {title} {{Microscopic Elasticity of
  Complex Systems}},\ }in\ \href {https://doi.org/10.1007/3-540-35284-8_12}
  {\emph {\bibinfo {booktitle} {Computer Simulations in Condensed Matter
  Systems: From Materials to Chemical Biology Volume 2}}}\ (\bibinfo
  {publisher} {Springer Berlin Heidelberg},\ \bibinfo {year} {2006})\ pp.\
  \bibinfo {pages} {287--307}\BibitemShut {NoStop}\bibitem [{\citenamefont {Lema{\^{\i}}tre}\ and\ \citenamefont
  {Maloney}(2006)}]{Lemaitre2006}\BibitemOpen
  \bibfield  {author} {\bibinfo {author} {\bibfnamefont {A.}~\bibnamefont
  {Lema{\^{\i}}tre}}\ and\ \bibinfo {author} {\bibfnamefont {C.}~\bibnamefont
  {Maloney}},\ }\bibfield  {title} {\bibinfo {title} {{Sum Rules for the
  Quasi-Static and Visco-Elastic Response of Disordered Solids at Zero
  Temperature}},\ }\href {https://doi.org/10.1007/s10955-005-9015-5} {\bibfield
   {journal} {\bibinfo  {journal} {Journal of Statistical Physics}\ }\textbf
  {\bibinfo {volume} {123}},\ \bibinfo {pages} {415} (\bibinfo {year}
  {2006})}\BibitemShut {NoStop}\bibitem [{\citenamefont {Zaccone}\ and\ \citenamefont
  {Scossa-Romano}(2011)}]{Zaccone2011}\BibitemOpen
  \bibfield  {author} {\bibinfo {author} {\bibfnamefont {A.}~\bibnamefont
  {Zaccone}}\ and\ \bibinfo {author} {\bibfnamefont {E.}~\bibnamefont
  {Scossa-Romano}},\ }\bibfield  {title} {\bibinfo {title} {Approximate
  analytical description of the nonaffine response of amorphous solids},\
  }\href {https://doi.org/10.1103/physrevb.83.184205} {\bibfield  {journal}
  {\bibinfo  {journal} {Physical Review B}\ }\textbf {\bibinfo {volume} {83}},\
  \bibinfo {pages} {184205} (\bibinfo {year} {2011})}\BibitemShut {NoStop}\bibitem [{\citenamefont {Liu}\ \emph {et~al.}(2011)\citenamefont {Liu},
  \citenamefont {Nagel}, \citenamefont {van Saarloos},\ and\ \citenamefont
  {Wyart}}]{Liu2011}\BibitemOpen
  \bibfield  {author} {\bibinfo {author} {\bibfnamefont {A.~J.}\ \bibnamefont
  {Liu}}, \bibinfo {author} {\bibfnamefont {S.~R.}\ \bibnamefont {Nagel}},
  \bibinfo {author} {\bibfnamefont {W.}~\bibnamefont {van Saarloos}},\ and\
  \bibinfo {author} {\bibfnamefont {M.}~\bibnamefont {Wyart}},\ }\bibfield
  {title} {\bibinfo {title} {The jamming scenario{\textemdash}an introduction
  and outlook},\ }in\ \href
  {https://doi.org/10.1093/acprof:oso/9780199691470.003.0009} {\emph {\bibinfo
  {booktitle} {Dynamical Heterogeneities in Glasses, Colloids, and Granular
  Media}}}\ (\bibinfo  {publisher} {Oxford University Press},\ \bibinfo {year}
  {2011})\ pp.\ \bibinfo {pages} {298--340}\BibitemShut {NoStop}\bibitem [{Note1()}]{Note1}\BibitemOpen
  \bibinfo {note} {This expression is taken at momentum $q = 0$ and the inverse
  $S^{-1}$ is computed in the orthogonal complement of the kernel of the
  matrix}\BibitemShut {NoStop}\bibitem [{\citenamefont {Avron}(1998)}]{Avron1998}\BibitemOpen
  \bibfield  {author} {\bibinfo {author} {\bibfnamefont {J.~E.}\ \bibnamefont
  {Avron}},\ }\href {https://doi.org/10.1023/a:1023084404080} {\bibfield
  {journal} {\bibinfo  {journal} {Journal of Statistical Physics}\ }\textbf
  {\bibinfo {volume} {92}},\ \bibinfo {pages} {543} (\bibinfo {year}
  {1998})}\BibitemShut {NoStop}\bibitem [{Note2()}]{Note2}\BibitemOpen
  \bibinfo {note} {In Eq.~\protect \textup {\hbox {\mathsurround \z@ \protect
  \normalfont (\ignorespaces \ref {consequences_duality_elastic_tensor}\unskip
  \@@italiccorr )}}, half of the indices of $c_{i \mu j \nu }$ are transformed
  by $\protect \mathcal {O}$ and the other half by $R$. In writing Eq.~\protect
  \textup {\hbox {\mathsurround \z@ \protect \normalfont (\ignorespaces \ref
  {hooke}\unskip \@@italiccorr )}}, we neglected the distinction between
  reference space (describing the undeformed elastic medium) and target space
  (describing the deformed medium), see Refs.~\cite
  {Truesdell1960,Truesdell2004,Ogden1997,Lubensky2002}. To emphasize this
  distinction, we use Latin indices for reference space coordinates $x_i$ and
  Greek indices for target space coordinates $X_\mu $. The displacement
  gradient $\epsilon _{i \mu } = \partial u_\mu /\partial x_i$ and the stress
  $\sigma _{i \mu }$ are now objects with mixed indices (mixed/two-point
  tensors) and Hooke's law reads $\sigma _{i \mu } = c_{i \mu j \nu } \epsilon
  _{j \nu }$ in contrast with Eq.~\protect \textup {\hbox {\mathsurround \z@
  \protect \normalfont (\ignorespaces \ref {hooke}\unskip \@@italiccorr )}}.
  This suggests that the elastic tensor $c_{i \mu j \nu }$ should not be
  restricted to transform according to Eq.~\protect \textup {\hbox
  {\mathsurround \z@ \protect \normalfont (\ignorespaces \ref
  {tensor_transformation_rule}\unskip \@@italiccorr )}}. Instead, different
  matrices can act on the reference and target spaces. These considerations
  also lead to the less restrictive form in Eq.~\protect \textup {\hbox
  {\mathsurround \z@ \protect \normalfont (\ignorespaces \ref
  {consequences_duality_elastic_tensor}\unskip \@@italiccorr )}}.}\BibitemShut
  {Stop}\bibitem [{\citenamefont {Guest}\ and\ \citenamefont
  {Hutchinson}(2003)}]{Guest2003}\BibitemOpen
  \bibfield  {author} {\bibinfo {author} {\bibfnamefont {S.}~\bibnamefont
  {Guest}}\ and\ \bibinfo {author} {\bibfnamefont {J.~W.}\ \bibnamefont
  {Hutchinson}},\ }\bibfield  {title} {\bibinfo {title} {{On the determinacy of
  repetitive structures}},\ }\href
  {https://doi.org/10.1016/s0022-5096(02)00107-2} {\bibfield  {journal}
  {\bibinfo  {journal} {Journal of the Mechanics and Physics of Solids}\
  }\textbf {\bibinfo {volume} {51}},\ \bibinfo {pages} {383} (\bibinfo {year}
  {2003})}\BibitemShut {NoStop}\bibitem [{\citenamefont {Hutchinson}\ and\ \citenamefont
  {Fleck}(2006)}]{Hutchinson2006}\BibitemOpen
  \bibfield  {author} {\bibinfo {author} {\bibfnamefont {R.}~\bibnamefont
  {Hutchinson}}\ and\ \bibinfo {author} {\bibfnamefont {N.}~\bibnamefont
  {Fleck}},\ }\bibfield  {title} {\bibinfo {title} {{The structural performance
  of the periodic truss}},\ }\href {https://doi.org/10.1016/j.jmps.2005.10.008}
  {\bibfield  {journal} {\bibinfo  {journal} {Journal of the Mechanics and
  Physics of Solids}\ }\textbf {\bibinfo {volume} {54}},\ \bibinfo {pages}
  {756} (\bibinfo {year} {2006})}\BibitemShut {NoStop}\bibitem [{\citenamefont {Paulose}\ \emph
  {et~al.}(2015{\natexlab{b}})\citenamefont {Paulose}, \citenamefont
  {ge~Chen},\ and\ \citenamefont {Vitelli}}]{Paulose2015a}\BibitemOpen
  \bibfield  {author} {\bibinfo {author} {\bibfnamefont {J.}~\bibnamefont
  {Paulose}}, \bibinfo {author} {\bibfnamefont {B.~G.}\ \bibnamefont
  {ge~Chen}},\ and\ \bibinfo {author} {\bibfnamefont {V.}~\bibnamefont
  {Vitelli}},\ }\bibfield  {title} {\bibinfo {title} {{Topological modes bound
  to dislocations in mechanical metamaterials}},\ }\href
  {https://doi.org/10.1038/nphys3185} {\bibfield  {journal} {\bibinfo
  {journal} {Nature Physics}\ }\textbf {\bibinfo {volume} {11}},\ \bibinfo
  {pages} {153} (\bibinfo {year} {2015}{\natexlab{b}})}\BibitemShut {NoStop}\bibitem [{\citenamefont {Truesdell}\ and\ \citenamefont
  {Noll}(2004)}]{Truesdell2004}\BibitemOpen
  \bibfield  {author} {\bibinfo {author} {\bibfnamefont {C.}~\bibnamefont
  {Truesdell}}\ and\ \bibinfo {author} {\bibfnamefont {W.}~\bibnamefont
  {Noll}},\ }\href {https://doi.org/10.1007/978-3-662-10388-3} {\emph {\bibinfo
  {title} {The Non-Linear Field Theories of Mechanics}}},\ edited by\ \bibinfo
  {editor} {\bibfnamefont {S.~S.}\ \bibnamefont {Antman}}\ (\bibinfo
  {publisher} {Springer Berlin Heidelberg},\ \bibinfo {year}
  {2004})\BibitemShut {NoStop}\bibitem [{\citenamefont {Ogden}(1997)}]{Ogden1997}\BibitemOpen
  \bibfield  {author} {\bibinfo {author} {\bibfnamefont {R.}~\bibnamefont
  {Ogden}},\ }\href {https://books.google.com/books?id=2u7wCaojfbEC} {\emph
  {\bibinfo {title} {Non-linear Elastic Deformations}}},\ Dover Civil and
  Mechanical Engineering\ (\bibinfo  {publisher} {Dover Publications},\
  \bibinfo {year} {1997})\BibitemShut {NoStop}\bibitem [{\citenamefont {Lubensky}\ \emph {et~al.}(2002)\citenamefont
  {Lubensky}, \citenamefont {Mukhopadhyay}, \citenamefont {Radzihovsky},\ and\
  \citenamefont {Xing}}]{Lubensky2002}\BibitemOpen
  \bibfield  {author} {\bibinfo {author} {\bibfnamefont {T.~C.}\ \bibnamefont
  {Lubensky}}, \bibinfo {author} {\bibfnamefont {R.}~\bibnamefont
  {Mukhopadhyay}}, \bibinfo {author} {\bibfnamefont {L.}~\bibnamefont
  {Radzihovsky}},\ and\ \bibinfo {author} {\bibfnamefont {X.}~\bibnamefont
  {Xing}},\ }\bibfield  {title} {\bibinfo {title} {Symmetries and elasticity of
  nematic gels},\ }\href {https://doi.org/10.1103/physreve.66.011702}
  {\bibfield  {journal} {\bibinfo  {journal} {Physical Review E}\ }\textbf
  {\bibinfo {volume} {66}},\ \bibinfo {pages} {011702} (\bibinfo {year}
  {2002})}\BibitemShut {NoStop}\bibitem [{GAP()}]{GAP}\BibitemOpen
  GAP,\ \href@noop {} {\bibinfo {title} {{GAP} {\textendash} {G}roups,
  {A}lgorithms, and {P}rogramming, {V}ersion 4.8.8}},\ \bibinfo {howpublished}
  {\href {https://www.gap-system.org} {\texttt{https://www.gap-system.org}}}
  (\bibinfo {year} {2017})\BibitemShut {NoStop}\bibitem [{\citenamefont {Felsch}\ and\ \citenamefont
  {G{\"a}hler}(2018)}]{CrystCat}\BibitemOpen
  \bibfield  {author} {\bibinfo {author} {\bibfnamefont {V.}~\bibnamefont
  {Felsch}}\ and\ \bibinfo {author} {\bibfnamefont {F.}~\bibnamefont
  {G{\"a}hler}},\ }\href
  {https://www.math.uni-bielefeld.de/~gaehler/gap45/packages.php} {\bibinfo
  {title} {{CrystCat}, the crystallographic groups catalog, {V}ersion 1.1.8}}
  (\bibinfo {year} {2018})\BibitemShut {NoStop}\bibitem [{\citenamefont {Lyubarskii}(1960)}]{Lyubarskii1960}\BibitemOpen
  \bibfield  {author} {\bibinfo {author} {\bibfnamefont {G.~Y.}\ \bibnamefont
  {Lyubarskii}},\ }\href@noop {} {\emph {\bibinfo {title} {The application of
  group theory in physics}}}\ (\bibinfo  {publisher} {Pergamon Press},\
  \bibinfo {year} {1960})\BibitemShut {NoStop}\bibitem [{\citenamefont {Yang}\ \emph {et~al.}(1994)\citenamefont {Yang},
  \citenamefont {hua Ding}, \citenamefont {Hu},\ and\ \citenamefont
  {Wang}}]{Yang1994}\BibitemOpen
  \bibfield  {author} {\bibinfo {author} {\bibfnamefont {W.}~\bibnamefont
  {Yang}}, \bibinfo {author} {\bibfnamefont {D.}~\bibnamefont {hua Ding}},
  \bibinfo {author} {\bibfnamefont {C.}~\bibnamefont {Hu}},\ and\ \bibinfo
  {author} {\bibfnamefont {R.}~\bibnamefont {Wang}},\ }\bibfield  {title}
  {\bibinfo {title} {{Group-theoretical derivation of the numbers of
  independent physical constants of quasicrystals}},\ }\href
  {https://doi.org/10.1103/physrevb.49.12656} {\bibfield  {journal} {\bibinfo
  {journal} {Physical Review B}\ }\textbf {\bibinfo {volume} {49}},\ \bibinfo
  {pages} {12656} (\bibinfo {year} {1994})}\BibitemShut {NoStop}\bibitem [{\citenamefont {Tinkham}(1992)}]{Tinkham1992}\BibitemOpen
  \bibfield  {author} {\bibinfo {author} {\bibfnamefont {M.}~\bibnamefont
  {Tinkham}},\ }\href@noop {} {\emph {\bibinfo {title} {Group Theory and
  Quantum Mechanics}}}\ (\bibinfo  {publisher} {Dover Publications},\ \bibinfo
  {year} {1992})\BibitemShut {NoStop}\bibitem [{\citenamefont {Dresselhaus}\ \emph {et~al.}(2008)\citenamefont
  {Dresselhaus}, \citenamefont {Dresselhaus},\ and\ \citenamefont
  {Jorio}}]{Dresselhaus2008}\BibitemOpen
  \bibfield  {author} {\bibinfo {author} {\bibfnamefont {M.~S.}\ \bibnamefont
  {Dresselhaus}}, \bibinfo {author} {\bibfnamefont {G.}~\bibnamefont
  {Dresselhaus}},\ and\ \bibinfo {author} {\bibfnamefont {A.}~\bibnamefont
  {Jorio}},\ }\href@noop {} {\emph {\bibinfo {title} {{Group Theory:
  Application to the Physics of Condensed Matter}}}}\ (\bibinfo  {publisher}
  {Springer},\ \bibinfo {year} {2008})\BibitemShut {NoStop}\bibitem [{\citenamefont {McWeeny}\ and\ \citenamefont
  {Jones}(1963)}]{McWeeny1963}\BibitemOpen
  \bibfield  {author} {\bibinfo {author} {\bibfnamefont {R.}~\bibnamefont
  {McWeeny}}\ and\ \bibinfo {author} {\bibfnamefont {H.}~\bibnamefont
  {Jones}},\ }\href@noop {} {\emph {\bibinfo {title} {{Symmetry: An
  Introduction to Group Theory and Its Applications}}}}\ (\bibinfo  {publisher}
  {Pergamon Press},\ \bibinfo {year} {1963})\BibitemShut {NoStop}\bibitem [{\citenamefont {DiDonna}\ and\ \citenamefont
  {Lubensky}(2005)}]{DiDonna2005}\BibitemOpen
  \bibfield  {author} {\bibinfo {author} {\bibfnamefont {B.~A.}\ \bibnamefont
  {DiDonna}}\ and\ \bibinfo {author} {\bibfnamefont {T.~C.}\ \bibnamefont
  {Lubensky}},\ }\bibfield  {title} {\bibinfo {title} {Nonaffine correlations
  in random elastic media},\ }\href
  {https://doi.org/10.1103/physreve.72.066619} {\bibfield  {journal} {\bibinfo
  {journal} {Physical Review E}\ }\textbf {\bibinfo {volume} {72}},\ \bibinfo
  {pages} {066619} (\bibinfo {year} {2005})}\BibitemShut {NoStop}\bibitem [{\citenamefont {Maloney}\ and\ \citenamefont
  {Lema{\^{\i}}tre}(2006)}]{Maloney2006}\BibitemOpen
  \bibfield  {author} {\bibinfo {author} {\bibfnamefont {C.~E.}\ \bibnamefont
  {Maloney}}\ and\ \bibinfo {author} {\bibfnamefont {A.}~\bibnamefont
  {Lema{\^{\i}}tre}},\ }\bibfield  {title} {\bibinfo {title} {{Amorphous
  systems in athermal, quasistatic shear}},\ }\href
  {https://doi.org/10.1103/physreve.74.016118} {\bibfield  {journal} {\bibinfo
  {journal} {Physical Review E}\ }\textbf {\bibinfo {volume} {74}},\ \bibinfo
  {pages} {016118} (\bibinfo {year} {2006})}\BibitemShut {NoStop}\bibitem [{\citenamefont {Rinaldi}\ and\ \citenamefont
  {Brenner}(2002)}]{Rinaldi2002}\BibitemOpen
  \bibfield  {author} {\bibinfo {author} {\bibfnamefont {C.}~\bibnamefont
  {Rinaldi}}\ and\ \bibinfo {author} {\bibfnamefont {H.}~\bibnamefont
  {Brenner}},\ }\bibfield  {title} {\bibinfo {title} {{Body versus surface
  forces in continuum mechanics: Is the Maxwell stress tensor a physically
  objective Cauchy stress?}},\ }\href
  {https://doi.org/10.1103/physreve.65.036615} {\bibfield  {journal} {\bibinfo
  {journal} {Physical Review E}\ }\textbf {\bibinfo {volume} {65}},\ \bibinfo
  {pages} {036615} (\bibinfo {year} {2002})}\BibitemShut {NoStop}\bibitem [{\citenamefont {Israelachvili}(2011)}]{Israelachvili2011}\BibitemOpen
  \bibfield  {author} {\bibinfo {author} {\bibfnamefont {J.~N.}\ \bibnamefont
  {Israelachvili}},\ }\href@noop {} {\emph {\bibinfo {title} {Intermolecular
  and surface forces}}}\ (\bibinfo  {publisher} {Academic press},\ \bibinfo
  {year} {2011})\BibitemShut {NoStop}\bibitem [{\citenamefont {Irving}\ and\ \citenamefont
  {Kirkwood}(1950)}]{Irving1950}\BibitemOpen
  \bibfield  {author} {\bibinfo {author} {\bibfnamefont {J.~H.}\ \bibnamefont
  {Irving}}\ and\ \bibinfo {author} {\bibfnamefont {J.~G.}\ \bibnamefont
  {Kirkwood}},\ }\bibfield  {title} {\bibinfo {title} {{The Statistical
  Mechanical Theory of Transport Processes. {IV}. The Equations of
  Hydrodynamics}},\ }\href {https://doi.org/10.1063/1.1747782} {\bibfield
  {journal} {\bibinfo  {journal} {The Journal of Chemical Physics}\ }\textbf
  {\bibinfo {volume} {18}},\ \bibinfo {pages} {817} (\bibinfo {year}
  {1950})}\BibitemShut {NoStop}\bibitem [{\citenamefont {Schofield}\ and\ \citenamefont
  {Henderson}(1982)}]{Schofield1982}\BibitemOpen
  \bibfield  {author} {\bibinfo {author} {\bibfnamefont {P.}~\bibnamefont
  {Schofield}}\ and\ \bibinfo {author} {\bibfnamefont {J.~R.}\ \bibnamefont
  {Henderson}},\ }\bibfield  {title} {\bibinfo {title} {{Statistical Mechanics
  of Inhomogeneous Fluids}},\ }\href {https://doi.org/10.1098/rspa.1982.0015}
  {\bibfield  {journal} {\bibinfo  {journal} {Proceedings of the Royal Society
  A: Mathematical, Physical and Engineering Sciences}\ }\textbf {\bibinfo
  {volume} {379}},\ \bibinfo {pages} {231} (\bibinfo {year}
  {1982})}\BibitemShut {NoStop}\bibitem [{\citenamefont {Wajnryb}\ \emph {et~al.}(1995)\citenamefont
  {Wajnryb}, \citenamefont {Altenberger},\ and\ \citenamefont
  {Dahler}}]{Wajnryb1995}\BibitemOpen
  \bibfield  {author} {\bibinfo {author} {\bibfnamefont {E.}~\bibnamefont
  {Wajnryb}}, \bibinfo {author} {\bibfnamefont {A.~R.}\ \bibnamefont
  {Altenberger}},\ and\ \bibinfo {author} {\bibfnamefont {J.~S.}\ \bibnamefont
  {Dahler}},\ }\bibfield  {title} {\bibinfo {title} {{Uniqueness of the
  microscopic stress tensor}},\ }\href {https://doi.org/10.1063/1.469942}
  {\bibfield  {journal} {\bibinfo  {journal} {The Journal of Chemical Physics}\
  }\textbf {\bibinfo {volume} {103}},\ \bibinfo {pages} {9782} (\bibinfo {year}
  {1995})}\BibitemShut {NoStop}\bibitem [{\citenamefont {Goldhirsch}(2010)}]{Goldhirsch2010}\BibitemOpen
  \bibfield  {author} {\bibinfo {author} {\bibfnamefont {I.}~\bibnamefont
  {Goldhirsch}},\ }\bibfield  {title} {\bibinfo {title} {{Stress, stress
  asymmetry and couple stress: from discrete particles to continuous fields}},\
  }\href {https://doi.org/10.1007/s10035-010-0181-z} {\bibfield  {journal}
  {\bibinfo  {journal} {Granular Matter}\ }\textbf {\bibinfo {volume} {12}},\
  \bibinfo {pages} {239} (\bibinfo {year} {2010})}\BibitemShut {NoStop}\bibitem [{\citenamefont {Rao}\ and\ \citenamefont {Bradlyn}(2020)}]{Rao2020}\BibitemOpen
  \bibfield  {author} {\bibinfo {author} {\bibfnamefont {P.}~\bibnamefont
  {Rao}}\ and\ \bibinfo {author} {\bibfnamefont {B.}~\bibnamefont {Bradlyn}},\
  }\bibfield  {title} {\bibinfo {title} {Hall viscosity in quantum systems with
  discrete symmetry: Point group and lattice anisotropy},\ }\href
  {https://doi.org/10.1103/physrevx.10.021005} {\bibfield  {journal} {\bibinfo
  {journal} {Physical Review X}\ }\textbf {\bibinfo {volume} {10}},\ \bibinfo
  {pages} {021005} (\bibinfo {year} {2020})}\BibitemShut {NoStop}\end{thebibliography}
\end{document}